\documentclass[%
  aps,pra,reprint,
  floatfix,a4paper,
  amsmath,amssymb,
  showkeys,nopreprintnumbers,
  showpacs,footinbib,
  superscriptaddress
  ,longbibliography,noeprint,nopreprintnumbers,
  twocolumn
  ]{revtex4-1}
\usepackage[T1]{fontenc}
\usepackage{graphicx}
\usepackage{bm}
\usepackage{comment}
\usepackage{MnSymbol} 
\usepackage{relsize} 
\usepackage{hyperref}
\usepackage[usenames,dvipsnames,svgnames,table]{xcolor}
\usepackage[percent]{overpic} 
\usepackage{braket}
\usepackage{textcase}
\usepackage[labelfont=bf,labelsep=period,justification=raggedright]{caption} 

 \newcommand{\mysection}[1]%
  {\smallskip\noindent\textbf{\textsc{#1}}\newline}
 \newcommand{\mysubsection}[1]{\noindent\textbf{#1}}
 \renewcommand{\figurename}{Figure} 

\newcommand{\vi}{\ensuremath{\vec{\imath}}}
\newcommand{\vf}{\ensuremath{\vec{f}}}

\newcommand{\calA}{{\cal A}} 
\newcommand{\calF}{{\cal F}} 
\newcommand{\partfgge}{{\ensuremath{\cal Z}}} 
\newcommand{\calP}{{\cal P}} 
\newcommand{\calW}{{\cal W}} 
\newcommand{\calWFW}{\ensuremath{{\cal W}_{\mathrm{FW}}}} 
\newcommand{\calWBW}{\ensuremath{{\cal W}_{\mathrm{BW}}}} 
\newcommand{\PFW}{\calP_{\mathrm{FW}}} 
\newcommand{\PBW}{\calP_{\mathrm{BW}}} 
\newcommand{\DFW}{\ensuremath{{\cal D}_{\mathrm{FW}}}} 
\newcommand{\DBW}{\ensuremath{{\cal D}_{\mathrm{BW}}}} 
\newcommand{\PFWm}{\PFW^{(m)}} 
\newcommand{\PBWm}{\PBW^{(m)}}
\newcommand{\pFW}{\ensuremath{P_{\mathrm{FW}}}} 
\newcommand{\pBW}{\ensuremath{P_{\mathrm{BW}}}}
\newcommand{\expval}[1]{\ensuremath{\langle{#1}\rangle}} 
\newcommand{\eval}[1]{\ensuremath{\langle{#1}\rangle}}
\newcommand{\deval}[1]{\ensuremath{\left\llangle{#1}\right\rrangle}} 
\DeclareMathOperator{\Tr}{Tr}

\newcommand{\U}{\ensuremath{\hat{U}}}
\newcommand{\tfin}{\ensuremath{\tau}} 
\newcommand{\Ham}{\ensuremath{\hat{H}}}
\newcommand{\Hini}{\ensuremath{\Ham}} 
\newcommand{\Hfin}{\ensuremath{\Ham'}}
\newcommand{\Eini}{\ensuremath{{\cal E}_\mathrm{ini}}} 
\newcommand{\Efin}{\ensuremath{{\cal E}_\mathrm{fin}}}

\newcommand{\rG}{\ensuremath{\hat{\rho}_{\textsc{gge}}}} 

\newcommand{\chargeQ}{M} 
\newcommand{\chargeQkini}{{\cal\chargeQ}_{k,\mathrm{ini}}}
\newcommand{\chargeQkfin}{{\cal\chargeQ}_{k,\mathrm{fin}}}
\newcommand{\chargeQlini}{{\cal\chargeQ}_{l,\mathrm{ini}}}
\newcommand{\chargeQlfin}{{\cal\chargeQ}_{l,\mathrm{fin}}}
\newcommand{\numcons}{\ensuremath{{\cal N}}} 
\newcommand{\numprot}{\ensuremath{{\cal N}_\mathrm{prot}}} 
\newcommand{\hQ}{\ensuremath{\hat{\chargeQ}}}
\newcommand{\hQm}{{\hat\chargeQ}_m}

\newcommand{\vecbeta}{\ensuremath{\vec{\beta}}}
\newcommand{\betaQ}{\ensuremath{\beta_{\chargeQ}}}
\newcommand{\omegaat}{\ensuremath{\omega_{\mathrm{at}}}}
\newcommand{\omegacom}{\ensuremath{\omega_{\textsc{com}}}}
\newcommand{\enUnit}{\ensuremath{\varepsilon_0}} 
\newcommand{\omegarsb}{\ensuremath{\Omega_{\mathrm{rsb}}}}
\newcommand{\omegabsb}{\ensuremath{\Omega_{\mathrm{bsb}}}}
\newcommand{\omegarsbbsb}{\ensuremath{\Omega_{\mathrm{bsb (rsb)}}}}
\newcommand{\hatb}{\ensuremath{\hat{b}}}
\newcommand{\hatJ}{\ensuremath{\hat{J}}}
\newcommand{\lambdaDicke}{\ensuremath{g}}
\newcommand{\gini}{\lambdaDicke_{\mathrm{ini}}}
\newcommand{\gint}{\lambdaDicke_{\mathrm{int}}}
\newcommand{\gfin}{\lambdaDicke_{\mathrm{fin}}}

\begin{document}

\def\mytitle{Revealing missing charges with generalised quantum fluctuation relations}
\title{\mytitle} 

\author{J. Mur-Petit}
\email[Electronic address: ]{jordi.murpetit@physics.ox.ac.uk}
\affiliation{Clarendon Laboratory, University of Oxford, 
	Oxford OX1 3PU, United Kingdom}
\author{A. Rela\~no}
\email[Electronic address: ]{armando.relano@fis.ucm.es}
\affiliation{Departamento de F\'{\i}sica Aplicada I and GISC, Universidad
Complutense de Madrid, Av.~Complutense s/n, 28040 Madrid, Spain}
\author{R. A. Molina}
\affiliation{Instituto de Estructura de la Materia, IEM-CSIC, Serrano 123, 28006 Madrid, Spain}
\author{D. Jaksch}
\affiliation{Clarendon Laboratory, University of Oxford, 
	Oxford OX1 3PU, United Kingdom}
\affiliation{Centre for Quantum Technologies, National University of Singapore,
	117543 Singapore}

\begin{abstract} 
The non-equilibrium dynamics of quantum many-body systems is one of the most fascinating problems in physics.
Open questions range from how they relax to equilibrium, to how to extract useful work from them.
A critical point lies in assessing whether a system has conserved quantities (or `charges'), as these can drastically influence its dynamics.
Here, we propose a general protocol to reveal the existence of charges based on a set of exact relations between out-of-equilibrium fluctuations and equilibrium properties of a quantum system.
We apply these generalised quantum fluctuation relations to a driven quantum simulator, demonstrating their relevance to obtain unbiased temperature estimates from non-equilibrium measurements.
Our findings will help guide research on the interplay of quantum and thermal fluctuations in quantum simulation, in studying the transition from integrability to chaos, and in the design of new quantum devices.
\end{abstract}


\date{\today}

\maketitle


\mysection{Introduction}
Einstein famously vouched for the enduring success of thermodynamics 
``within the framework of the applicability of its basic concepts'' stemming from the simplicity of its premises and breadth of its scope~\cite{Schilpp2000}.
From its birth as a practical science in the cradle of the industrial revolution~\cite{Carnot1824} to modelling thermal fluctuations in biological processes through fluctuation relations (FRs)~\cite{Liphardt2002,Ritort2008}, 
thermodynamics constitutes one of the most successful theories to understand Nature.
The increasing degree of control on meso- and nano-scopic systems has driven interest into the field of quantum thermodynamics to describe phenomena where both quantum effects and finite-size fluctuations are apparent~\cite{Esposito2009}. Important findings so far range from generalised Carnot bounds on the efficiency of quantum heat engines~\cite{Scully2003,
Abah2014,Rossnagel2014}, to quantum versions of the classical FRs --i.e., quantum fluctuation relations (QFRs)-- for processes starting in a canonical equilibrium state~\cite{Hanggi2015}.
According to the principles of quantum statistical mechanics, such a state is characterised by a single parameter, the inverse temperature $\beta$, which also plays a special role in the QFRs (see~\cite{Jarzynski2015commentary} and references therein).

The dynamics of an important number of quantum systems, however, eludes this approach. Integrable quantum systems, for instance, feature a large number of conserved quantities, or `charges', which effectively constrain the phase space that the system can explore in its dynamic evolution~\cite{Sutherland2004}. 
Notable models with charges include the one-dimensional Hubbard model~\cite{Essler2005}, the Dicke model~\cite{Dicke1954,Garraway2011}, and the super-symmetric $t$-$J$ model~\cite{Sutherland1975},
to name a few. %
The existence of charges leads sometimes to striking experimental observations such as the practically dissipationless dynamics of the quantum Newton's cradle~\cite{Kinoshita2006}, which stems from the (infinitely many) charges of the Lieb-Liniger model. 
On other occasions, however, their existence is far from obvious.
For instance, it is only recently that a whole set of quasi-local charges in XXZ model were discovered~\cite{Prosen2011,Prosen2013,Ilievski2017} as result of a discrepancy observed between numerical simulations of the model and estimates based on the Mazur bound~\cite{Heidrich-Meisner2003}.

It is a critical task of quantum many-body physics to develop methods to confidently ascertain whether a quantum system features such charges, especially when these may be difficult to measure directly.
Unfortunately, the Mazur bound is only sensitive to charges with a non-zero overlap with the current operator, which implies it cannot serve as a general witness to unveil all charges in a generic quantum many-body system. 
It is thus necessary to develop more general systematic methods to explore the existence of unknown conserved charges in quantum many-body systems.

Here, we introduce an approach to this problem based on a statistical analysis of arbitrary non-equilibrium measurements of the system of interest.
In short, we demonstrate that non-equilibrium measurements on a quantum many-body system are more sensitive to the existence of charges than equilibrium ones, and describe a protocol that exploits this sensitivity to reveal the existence of charges that restrict the dynamics in any degree of freedom of the system.
To this end, we first provide a theory that completely characterises the non-equilibrium fluctuations of quantum systems with conserved quantities.
Specifically, we present generalised versions of the Tasaki-Crooks relation (TCR)~\cite{Tasaki2000b} and the quantum Jarzynski equality (QJE)~\cite{Tasaki2000a,Kurchan2000,Yukawa2000}
suitable to describe fluctuations in systems with an arbitrary, and possibly variable, number of charges.
Then, we show how these results open the door to determining from experimental measurements the existence of hitherto unknown charges. We also discuss how this can improve the accuracy, e.g., of temperature measurements.
Finally, we illustrate our results with simulations of the Dicke model~\cite{Dicke1954,Garraway2011}, a well-known many-body model that features a single charge and which can be realised in current experimental platforms.

\begin{figure}[tbh]
\centering
 \includegraphics[width=\columnwidth]{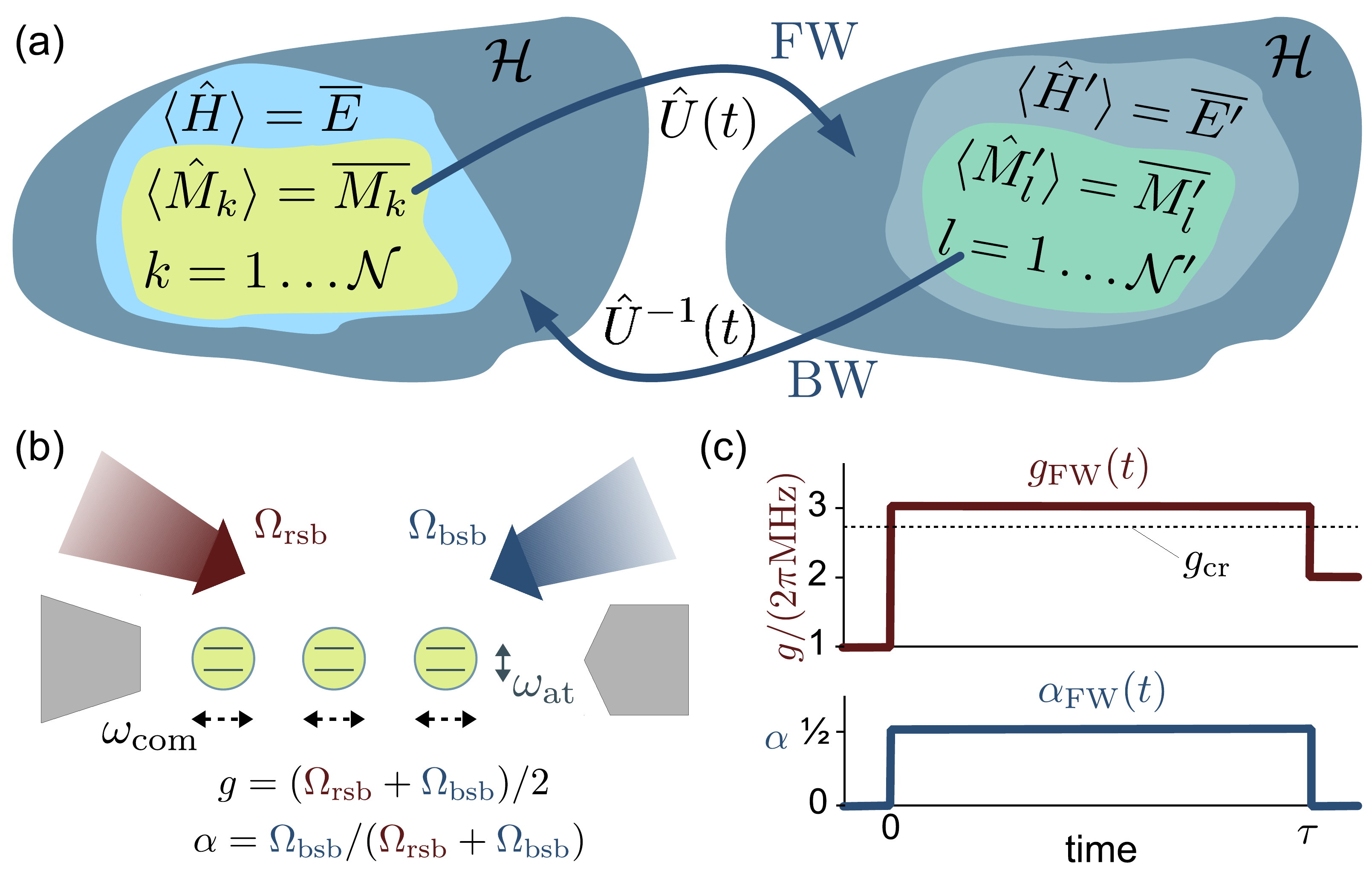}
\caption{
 \textbf{Sketch of the system and protocol.}
 (a) The dynamics of a generic quantum system with average energy $\expval{\Ham}=\overline{E}$ occurs within a restricted subspace (light blue area) of its full Hilbert space, $\cal H$ (dark blue). If additional conserved quantities exist, the dynamics is further restricted to a smaller subspace (yellow). An equilibrium state of such a system with charges is described by a generalised Gibbs ensemble density matrix, Eq.~(\ref{eq:rhoGGE}).
 Here, we consider two unitary processes $\U(t), \U^{-1}(t)$ that drive the system out of two such equilibrium states corresponding to Hamiltonians $\Ham$ and $\Ham'$, respectively.
 (b) Trapped ion setup: $N$ ions (circles) equally coupled to a phonon mode (black arrows) are illuminated by fields addressing the red and blue sidebands (wide red and blue arrows) with Rabi frequency $\omegarsb \:[\omegabsb]$.
 (c) Time dependence of the Dicke model parameters in the forward (`FW') protocol, with a variable wait time $\tfin$ between two quenches.
 }
\label{fig:sketches}
\end{figure}

\mysection{Results}
\mysubsection{Theoretical framework}
The quantum-statistical description of systems with charges can be reliably built on Jaynes' information-theory formulation of statistical mechanics~\cite{Jaynes1957a}. 
In this approach, a new statistical ensemble, the generalised Gibbs ensemble (GGE), has been proposed~\cite{Rigol2007} to incorporate the constraints on the known values of the charges to the equilibrium state via the maximum entropy principle (see also~\cite{Moreno-Cardoner2007}). 
In the GGE, the equilibrium state of a system with Hamiltonian $\Ham$ is given by a density matrix of the form
\begin{align}
 \rG
 &
 = \frac{1}{\partfgge} \exp \left( -\beta \Ham -\sum_{k=1}^{\numcons} \beta_k \hQ_k \right) 
 \:,
 \label{eq:rhoGGE}
\end{align}
where $\partfgge\equiv \partfgge(\vecbeta,\Hini,\{\hQ_k\})= \mathrm{Tr}[\exp(-\beta\Ham-\sum_k \beta_k \hQ_k)]$ is the partition function, and the operators associated with the charges, $\hQ_k$, satisfy $[\hQ_k, \Ham]=0$, for $k=1,\ldots,\numcons$, with $\numcons$ the number of charges of the system, see Fig.~\ref{fig:sketches}(a).
(Below, we will assume the charge operators commute with each other, which enables measuring them simultaneously.)
The generalised inverse temperatures, $\vecbeta=(\beta, \, \{ \beta_k \}_{k=1}^\numcons )$, are fixed by requiring that averages over $\rG$ reproduce the known average values of the energy, $\braket{\Ham}\equiv \Tr[\Ham \rG] = \overline{E}$, and charges, $\braket{\hQ_k} = \overline{\chargeQ}_k$. 
A crucial open question springing from Eq.~(\ref{eq:rhoGGE}) is the identification of all charges $\hQ_k$ relevant to the dynamics of the system. The usual approach consists in the study of equilibrium expectation values of certain observables. However, this approach suffers from a number of caveats and difficulties, like the very need of measuring a large number of observables, or the existence of particular observables that may not thermalise (see Supplementary Note 1 for more details). We show below that measurements in non-equilibrium processes are highly sensitive to the existence of charges, and how one can use them to reveal the presence of conserved quantities in the equilibrium state.



\mysubsection{Generalised QFRs}
We study the energy fluctuations of a system with charges by considering two processes (`forward' and `backward') that take the system away from initial equilibrium states, cf.\ Fig.~\ref{fig:sketches}(a).
Each process is a four-step protocol similar to the two-projective-measurements (TPM) protocol utilised to derive the standard QFRs~\cite{Talkner2007}.

In the ``forward'' (FW) process, the system is
(i) prepared in the equilibrium state corresponding to Hamiltonian $\Hini$. If this Hamiltonian features a number $\numcons$ of charges,
this state can be written in the form of Eq.~(\ref{eq:rhoGGE}) with $\numcons+1$ parameters $\vecbeta=\{ \beta, \beta_1, \ldots, \beta_\numcons \}$.
We build a basis of the Hilbert space with eigenvectors $\ket{\vi} = \ket{i_0, i_1, \ldots, i_{\numcons}, \eta }$, 
where the quantum number $i_0$ identifies the energy eigenvalue, $\Hini\ket{\vi} = E_{i_0} \ket{\vi}$,
and $i_k$ similarly labels the eigenvalues of $\hQ_k$ through 
$\hQ_k\ket{\vi} = \chargeQ_{k,i_k} \ket{\vi}$ ($k=1,\ldots,\numcons$); $\eta$ contains the additional quantum numbers required to fully determine a basis state.
After this preparation stage,
(ii) at time $t=0$ one performs simultaneous projective measurements of $\Ham$ and $\hQ_k$ on the system, obtaining definite values for its energy, $\Eini \in \{ E_i \}$, and the other observables, $\chargeQkini \in \{ \chargeQ_{k,i_k} \}$.
(iii) In the third step, the system is driven out of equilibrium by steering its Hamiltonian in a time-dependent process, $\Ham \mapsto \Ham(t)$, for times $0 < t < \tfin$. This defines a unitary time-evolution operator $\U(t)$ as the solution of $i\hbar\partial_t \U(t) = \Ham(t) \U(t)$, with $\U(0)=\mathbb{I}$, the identity operator on the system's Hilbert space $\cal H$.
Finally, (iv) at time $t=\tfin$, the system is projected on the eigenbasis of the instantaneous Hamiltonian, $\Hfin = \Ham(\tfin)$.
In general, the operators $\hQ_k$ will not commute with $\Ham(t)$ for $t>0$, and we denote
the set of charges that commute with $\Hfin$ as $\{ \hQ_l' \} \: (l=1, \ldots,\numcons' )$.
Assuming that these operators commute with each other, this second projective measurement provides the quantities $\Efin'$ and $\{ \chargeQlfin' \}$, each belonging to the spectrum of the corresponding operator, in full analogy to the situation at $t=0$.
Thus, at the end of a single realisation of the FW process, one has collected the dataset $\DFW = \{ \Eini, \{ \chargeQkini \} ; 
\Efin', \{ \chargeQlfin' \} \}$ associated to the parameters $\vecbeta$ of the initial state Eq.~(\ref{eq:rhoGGE}).

The complementary ``backward'' (BW) process starts by preparing the system in an equilibrium state of the Hamiltonian $\Hfin$, cf.\ Fig.~\ref{fig:sketches}a. In accordance with the preceding discussion, this state will be of the GGE form with $\numcons'+1$ parameters, $\vecbeta'=\{ \beta', \beta_1', \ldots, \beta_{\numcons'}' \}$.
At time $t=0$, the system is projected on the basis $\ket{\vf'}$ of simultaneous eigenvectors of $\Hfin$ and $\hQ_l'$, obtaining the values $\Eini'$ and
$\{ \chargeQlini' \}$ for the corresponding observables.
The system then evolves under the time-reversed protocol $\U^{-1}(t)$ for $0 < t < \tfin$, so that at time $\tfin$ its Hamiltonian is $\Hini$. A projective measurement on the final instantaneous eigenbasis provides values $\Efin$ and
$\{ \chargeQkfin \}$ for the energy and other observables.
A single realisation of the BW protocol thus gives a dataset 
$\DBW = \{ \Eini', \{ \chargeQlini' \} ; \Efin, \{ \chargeQkfin \} \}$
associated to the parameters $\vecbeta'$ of the BW initial state.

With the datasets $\DFW$ and $\DBW$ we build two (dimensionless) work-like quantities
\begin{align}
 \calWFW
 &\equiv
 \beta' \Efin' + \sum_{l=1}^{\numcons'} \beta_l' \chargeQlfin'
 - \Big[ \beta \Eini + \sum_{k=1}^{\numcons} \beta_k \chargeQkini \Big] ,
 \label{eq:gen-work-fw}
 \\
 \calWBW
 &\equiv
 \beta \Efin + \sum_{k=1}^{\numcons} \beta_k \chargeQkfin
 - \Big[ \beta' \Eini' + \sum_{l=1}^{\numcons'} \beta_l' \chargeQlini' \Big] .
 \label{eq:gen-work-bw}
\end{align}
Due to the projective nature of the measurements, both these quantities are stochastic variables, and their statistics can be described through probability distribution functions (PDFs), $\PFW$ and $\PBW$, associated respectively to the FW and BW processes.
We find that, although the initial states of the two processes are independent, and may feature different numbers of charges and generalised inverse temperatures, these PDFs are not independent, but obey the following relation (see Methods):
\begin{align} 
 \frac{ \PFW(\calW) }{ \PBW(-\calW) }
 e^{-\calW}
 & = 
\frac{\partfgge'}{\partfgge} \:,
 \label{eq:gen-tcr}
\end{align}
where $\partfgge' \equiv \partfgge(\vecbeta',\Hfin,\{ \hQ_l' \})$ is the equilibrium partition function of the initial state of the BW process.
By introducing (dimensionless) generalised free energy functions as 
$\calF = -\ln\partfgge$ and $\calF' = -\ln\partfgge'$, the  right hand side (r.h.s.) 
of Eq.~(\ref{eq:gen-tcr}) becomes
$\exp(\calW-\Delta \calF)$, with $\Delta \calF = \calF'-\calF$.
Eq.~(\ref{eq:gen-tcr}) is the generalisation of the TCR to systems with arbitrary numbers of charges associated to each equilibrium state, and to out-of-equilibrium processes that change the number of charges.

If we multiply both sides of Eq.~(\ref{eq:gen-tcr}) by $\PBW(-\calW)$ and integrate over $\calW$, we get
\begin{align} 
 \deval{e^{-\calW}}_{\mathrm{FW}}
 &\equiv 
 \int_{-\infty}^{\infty} \mathrm{d}\calW \: \PFW(\calW) \: e^{-\calW}
 = e^{-\Delta \calF} \:,
 \label{eq:gen-qje}
\end{align}
which is a generalisation of the QJE for systems with charges.
Here, we remark a qualitative difference between Eqs.~(\ref{eq:gen-tcr}) and~(\ref{eq:gen-qje}). While the TCR relates the outcomes of driving processes starting in two different initial states, the QJE applies to a single system driven out of an equilibrium characterised by parameters $\vecbeta$.
The apparent dependence on $\vecbeta'$ of Eq.~(\ref{eq:gen-qje}) in fact shows that 
this equation relates an initial equilibrium state to all possible GGE-like states of the final Hamiltonian: for each possible set of `final' equilibrium parameters $\vecbeta'$, the numerical values on the left hand side (l.h.s.) and the r.h.s.\ of Eq.~(\ref{eq:gen-qje}) will differ, but the equality will hold as long as the initial state is of the GGE form, Eq.~(\ref{eq:rhoGGE}). In this sense, the $\vecbeta'$ dependence in Eq.~(\ref{eq:gen-qje}) is irrelevant, and one can test its validity taking e.g. $\vecbeta'=\vecbeta$.
This agrees with the intuition that by driving the system out of equilibrium we can obtain information on its initial equilibrium state (i.e., $\vecbeta$),
but we cannot give physical meaning to the values of $\vecbeta'$ in Eq.~(\ref{eq:gen-qje}).
(The physics of Eqs.~(\ref{eq:gen-tcr})-(\ref{eq:gen-qje}) is further discussed in
Supplementary Note 2.)

More generally, consider the following scenario in which one of the constants of motion, say $\hQm$, commutes with the time-dependent Hamiltonian and with all the other charges at all times.
Then, the time-dependent Hamiltonian can be set in a block-diagonal form, with different blocks corresponding to the different eigenvalues of this operator, and the values of $\hQ_m$ measured at the start and end of the TPM protocol must be identical.
In this case, let us introduce a marginal generalised work $\calW_m$ by
$\calW_m
 \equiv \beta'\Efin + \sum_{l\neq m} \beta_l' \chargeQlfin'
 -( \beta \Eini + \sum_{k\neq m} \beta_k \chargeQkini)$.
The corresponding PDF,
$ \PFWm(\calW_m) $,
satisfies a marginal version of the generalised TCR (see Methods):
\begin{align}
 \frac{ \PFWm(\calW_m) }{ \PBWm(-\calW_m) }
 e^{-\calW_m} 
 = 
 e^{-\Delta\calF} \:.
 \label{eq:marginal-tcr}
\end{align}
We show below how this result can be used to check whether a particular observable does or does not change in a non-equilibrium process without measuring it.


\mysubsection{Revealing missing charges}
We now discuss how Eqs.~(\ref{eq:gen-tcr}), (\ref{eq:gen-qje}), and~(\ref{eq:marginal-tcr})
underlie novel strategies for two applications: (i) to reveal hidden conserved charges, and (ii) to check whether a particular observable of difficult experimental access does or does not change during a quantum non-equilibrium protocol.
Next, we will illustrate this in practice through numerical studies of a trapped-ion quantum simulator.

The key realisation is that the r.h.s.\ of Eqs.~(\ref{eq:gen-tcr}), (\ref{eq:gen-qje}), and~(\ref{eq:marginal-tcr}) is determined solely by equilibrium properties pertaining to the initial states of the FW and BW processes respectively, while the l.h.s.\ relates to measurement outcomes of non-equilibrium processes starting from those initial states, obtained via the TPM protocol.
We can then check the completeness of the set of known charges by running the TPM protocol with a number, $\numprot$, of different protocols $\U$ (corresponding, for instance, to different durations $\tfin$).
The experimental data corresponding to the $\numprot$ protocols provide different values for the l.h.s.\ of the generalised QJE, Eq.~(\ref{eq:gen-qje}).
As discussed above, these values depend physically on the $\numcons+1$ parameters $\vecbeta$ of the initial GGE state.
The fact that all these expressions must equal the same single value on the r.h.s.\ entails a set of $\numprot$ (nonlinear) equations for $\numcons+1$ unknowns. 
If we can find values for $\vecbeta$ that satisfy these equations, then we have accounted for all the charges required to describe the non-equilibrium dynamics of the system induced by $\U$; on the other hand, failure to find a satisfactory set of parameters points that more charges need to be included. A similar argument can be made based on the generalised TCR~(\ref{eq:gen-tcr}), to check whether all the charges in both FW and BW initial states have been accounted for. (Generally, the TCR depends on the $\numcons+\numcons'+2$ parameters $\{\vecbeta,\vecbeta'\}$ characterising the FW and BW initial states. In practice, the number of required experiments can be notably reduced, e.g., by putting the system in contact with the same bath at the start of both FW and BW processes, so that $\vecbeta=\vecbeta'$, in which case only $\numcons+1$ unknowns need to be determined.)
We note in addition that these completeness tests do not require prior knowledge of the inverse temperatures characterising the initial states; however, if we do have a reliable preliminary estimate of the inverse temperatures, this test allows to verify this estimate, or to conclude that the number of known charges is insufficient as soon as inconsistencies with Eqs.~(\ref{eq:gen-tcr})-(\ref{eq:gen-qje}) emerge.

As a complement, Eq.~(\ref{eq:marginal-tcr}) enables us to assess whether an operator that is known to commute with a Hamiltonian $\Ham$ does, or does not, change in a non-equilibrium procedure without measuring it, i.e., it enables to determine whether a non-equilibrium procedure transforms a charge into a dynamical variable.
To see this, let us imagine that we are interested to know whether a certain observable of difficult experimental access, $\hQ_m$, is or not perturbed by a certain class of non-equilibrium protocols.
To assess this, we perform a set of FW and BW protocols, excluding $\hQ_m$ from 
the TPM measurements, and calculate 
the marginal work $\calW_m$. With these values, we build the associated PDFs, $\PFWm$ and $\PBWm$.
If we can find values of $\vecbeta$ and $\vecbeta'$ without $\beta_m$ such that the r.h.s.\ of Eq.~(\ref{eq:marginal-tcr}) matches the data on the l.h.s., then the contributions of $\hQ_m$ to $\calF$ and $\calF'$ must cancel
one another in $\Delta\calF$, i.e., $\hQ_m$ has remained constant throughout the process. Otherwise, $\hQ_m$ cannot have the same value at the start and end of the protocol, i.e., $\hQ_m$ is a dynamical variable in the process.


\mysubsection{Application to a trapped-ion quantum simulator}
Our generalised QFRs are valid for arbitrary unitary non-equilibrium processes, $\U$, applied to quantum systems with conserved quantities~\cite{Kinoshita2006,Gring2012,Langen2015,Ronzheimer2013}.
In the following, we illustrate their implications in the context of a trapped ion experiment realisable with current technology~\cite{Benhelm2008pra,Harty2014,An2015,Kienzler2017}.
First we show that this system can be described by a Hamiltonian with a single charge. We then report numerical evidence showing how measurements of its work statistics in generic non-equilibrium protocols would violate the standard QJE and TCR ---
and how they agree with the predictions of our generalised QFRs.
The fact that this model features a single charge makes it an especially attractive test ground, as this makes experimental tests of our generalised QFRs far more accessible than other models that would in principle require measurements on an infinite number of charges, such as the XXZ model.

We consider $N$ $^{43}$Ca$^+$ ions in an ion trap~\cite{Benhelm2008pra,Harty2014}, see Fig.~\ref{fig:sketches}(b).
Each ion can be described as a two level system with internal states corresponding to two Zeeman levels within the ground $^2S_{1/2}$ electronic state, whose energy splitting, $\hbar\omegaat$, can be controlled by an external bias magnetic field~\cite{Benhelm2008pra,Harty2014}.
The motional state of the ions in the trap is characterised by $N-1$ collective modes in each direction~\cite{Haffner2008}.
Among these, we focus on the centre-of-mass (COM) mode, which couples identically to all ions and whose eigenfrequency, $\omegacom$, is of the order of the trap's oscillator frequency~\cite{Haffner2008}.
Internal and motional states can be coupled by light fields of frequency $\omega$ close to $\omega_{\pm} = \omegaat \pm\omegacom$, the blue (+) and red ($-$) motional sidebands.
The Hamiltonian describing the dynamics of this system can be written in the form (see Methods and~\cite{Emary2003prl,Emary2003pre,Relano2016,Kienzler2017})
\begin{align}
 H/\hbar
 &=
 \omegacom \hatb^{\dagger} \hatb 
 +\omegaat \hatJ_z
 + \frac{2\lambdaDicke}{\sqrt{N}} \Big[ 
 	(1-\alpha) \left( \hatJ_+ \hatb + \hatJ_- \hatb^{\dagger} \right)
 \nonumber \\
 &+\alpha \left( \hatJ_+ \hatb^{\dagger} + \hatJ_- \hatb \right)
 \Big]
 \label{eq:Dicke}
\end{align}
where $\hatb^\dagger$ and $\hatb$ are the operators creating and annihilating excitations in the COM mode, and $\hatJ_s$ ($s=z,+,-$) are Schwinger spin operators describing the internal state of all the ions, with $J=N/2$.
In Eq.~(\ref{eq:Dicke}) we have introduced $\lambdaDicke=(\omegarsb+\omegabsb)/2$ and $\alpha=\omegabsb/(\omegarsb+\omegabsb)$, with $\Omega_{\mathrm{bsb\,(rsb)}}$ the Rabi frequency characterising the coupling of internal and motional states through the first blue (red) motional sideband; these are functions of the light intensity at frequency
$\omega_{\pm}$ respectively~\cite{Haffner2008,Kienzler2017}.
The Hamiltonian Eq.~(\ref{eq:Dicke}) is exactly that of the Dicke model~\cite{Dicke1954,Garraway2011}.
For $\alpha=0$, it reduces to the Tavis-Cummings model, which is integrable and has an additional conserved quantity,
$\hQ = \hatJ + \hatJ_z + \hatb^\dagger \hatb$
(see Methods and 
Supplementary Note 3). Thus, the dynamics of this
system starting from an equilibrium state will be governed by our generalised QFRs.
This can be verified by extending the filtering method~\cite{Huber2008,An2015} used to verify the standard QJE~\cite{An2015}, to account for the internal structure of the ions in the Dicke model, so as to determine the initial and final energy values; indeed, as the spectrum of the Dicke model is non-degenerate, an energy measurement provides both $\cal E$ and $\cal \chargeQ$.
Additionally, the PDF of standard work can be obtained without projective measurements by utilising an ancilla qubit~\cite{Dorner2013,Mazzola2013,Batalhao2014}.

In order to assess the existence of charges, we drive the system out of equilibrium. For simplicity, we consider a series of sudden quenches in space $\{ g, \alpha \}$; experimentally, these quenches correspond to changes in the intensities of the lasers realising the sideband couplings on a timescale much shorter than $\omegacom^{-1}$.
We consider in particular the FW protocol $\{\gini,0\} \rightarrow \{\gint,1/2\} \rightarrow \{\gfin,0\}$, with the system remaining in the intermediate stage for a variable time $\tfin$, see Fig.\ \ref{fig:sketches}(c);
this duration plays the role of the parameter $\tfin$ characterising the $\numprot$ protocols in the procedure to reveal missing charges described above.

The choice $\alpha=0$ at $t \in \{0, \tfin\}$ ensures that $\hQ$ commutes with both $\Hini$ and $\Hfin$.
Hence, the initial equilibrium state of FW and BW processes will be of the GGE form with specific values for the inverse temperatures related to $\Ham$ and $\hQ$, which we label $\beta$ and $\betaQ$ (for concreteness, we analyse here the case that $\vecbeta'=\vecbeta$; see Supplementary Note 2 for a discussion on this choice).
However, in the intermediate stages $\alpha=1/2$, which implies that there are no charges during an important part of the process.
We will see that it is nevertheless possible to determine whether the system had a conserved quantity at the start of the process.
(Additional simulations in 
Supplementary Note 4 for a process ending in $\alpha=1/2$, i.e., where $\hQ$ and $\Hfin$ do not commute, support analogous conclusions, see Supplementary Fig.~1.)


\begin{figure*}[tbh]
\centering
\includegraphics[width=\textwidth]{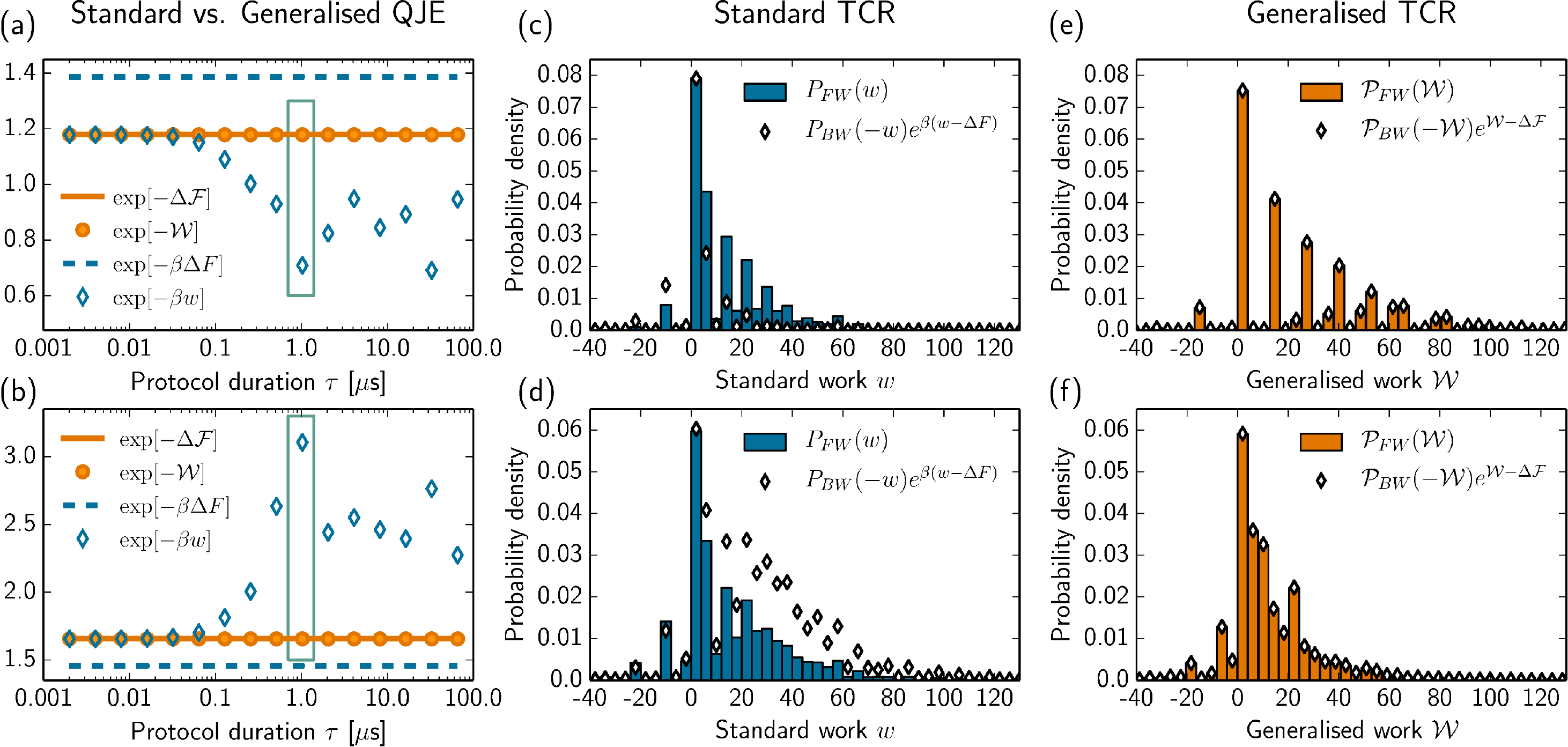}
\caption{%
 \textbf{Generalised quantum fluctuation relations.}
  (a) Plot of $\exp(-\Delta\calF)$ (solid orange line)
    and $\exp(-\beta\Delta F)$ (dashed blue line),
    compared with the exponentiated averages
    $\deval{\exp(-\calW)}$ (circles)
    and $\deval{\exp(-\beta w)}$ (open diamonds)
    for protocols with duration $\tfin \in [1~\mathrm{ns},\: 100~\mu\mathrm{s}]$
    that start in a GGE state given by $(\beta\enUnit=0.1, \: \betaQ\enUnit=0.3)$.
  (b) Same as (a) with $\betaQ\enUnit=-0.1$.  
  (c) Probability distribution function (PDF) of standard work for the FW process,
   $\pFW(w)$ (blue bars), and PDF for the BW process weighted according to the standard TCR,
   $\exp[\beta(w-\Delta F)] \pBW(-w)$, (diamonds) for the process with duration
   $\tfin=1.024~\mu$s [green box in (a)].
  (d) Same as (c) for the process indicated in (b).
  (e,f) PDF of generalised work for the FW process, $\PFW(\calW)$ (orange bars),
   and PDF of generalised work for the BW process weighted according to the generalised TCR
   $\exp(\calW-\Delta\calF) \PBW(-\calW)$, (diamonds) for the process indicated by
   a box in (a, b) respectively.
  Parameters for all simulations: $N=7$ ions,
  $\omegacom=3\enUnit$, $\omegaat=10\enUnit$,
  $\gini=2\enUnit$,
  $\gint=3\enUnit$, and
  $\gfin=\enUnit$,
  with $\enUnit=\hbar\times 2\pi$~MHz a typical value trapping energy scale in trapped-ion experiments~\cite{Benhelm2008pra,Harty2014,An2015,Kienzler2017}.
 }
\label{fig:results-all}
\end{figure*}


We show in Fig.~\ref{fig:results-all}(a) the average exponentiated work
performed on the system by a FW process as a function of duration $\tfin$.
We compare the results using standard work, $\deval{\exp(-\beta w)}_{\mathrm{FW}}$, to those with generalised work $\deval{\exp(-\calW)}_{\mathrm{FW}}$.
We see that the standard work average varies by up to 40\% for durations $\tfin\gtrsim 0.1~\mu$s.
This $\tfin$ dependence indicates that the standard work is no longer the relevant magnitude in non-equilibrium processes:
there is one (or more) missing charge(s) whose fluctuations need to be taken into account to describe the measurement outcomes.
In contrast to this, the average of generalised work including $\hQ$ remains constant for all $\tfin$; it thus follows from Eq.~(\ref{eq:gen-qje}) that this definition of $\calW$ includes all the relevant charges of the system.

Notably, the two averages agree with each other for short processes, $\tfin < 0.1~\mu$s. This reflects that, at short times, $\hQ$ remains approximately constant and a marginal TCR is expected to hold.
Importantly, however, both averages agree with the prediction of the generalised QFRs, i.e., $\exp(-\Delta\calF)$, while the expectation that excludes $\betaQ\hQ$ (dashed line) is off by 20\%.
In practice, this means that a na\"{\i}ve fit of the $\deval{\exp(-\beta w)}$ data to $\exp(-\beta\Delta F)$ would yield a biased value for $\beta$, i.e., a wrong estimate of the initial temperature of the system.
Generally, use of the standard QFRs will provide biased estimates if the system has charges.
Fig.~\ref{fig:results-all}(b) support analogous conclusions for the case $\betaQ<0$.

The relevance of charge fluctuations in the non-equilibrium dynamics becomes apparent in Figs.~\ref{fig:results-all}(c-f), which portray the statistics of standard and generalised work under driving protocols with $\tfin = 1~\mu$s.
In Figs.~\ref{fig:results-all}(c,d) we observe that scaling the PDF of standard work for the BW process by $\exp[\beta(w-\Delta F)]$ (diamonds) does not agree with the PDF of the FW process (bars), in contrast to the prediction of the standard TCR.
Such a disagreement is a strong indication of the existence of charges in a system.
On the other hand,  the prediction of the generalised TCR, Eq.~(\ref{eq:gen-tcr}), is satisfied with great accuracy for both $\betaQ>0$ [Fig.~\ref{fig:results-all}(e)] and $\betaQ<0$ [Fig.~\ref{fig:results-all}(f)].
In other words, the TCR equality is only fulfilled when all relevant charges are included in the calculations of $\calW$ and $\Delta\calF$, and a discrepancy between the measurable quantities $\PFW(\calW)$ and $\PBW(\calW) \exp(\calW-\Delta\calF)$ points to the existence of hidden charges that need to be included.


\mysection{Discussion}
%
Our numerical results in Fig.~\ref{fig:results-all}(a-d) highlight the limitations of the standard QFRs in dealing with systems with charges, in particular in order to extract equilibrium properties by means of non-equilibrium measurements~\cite{Dorner2013,Mazzola2013,Batalhao2014,Johnson2016thermo,Streif2016}.
Our generalised QFRs, Eqs.~(\ref{eq:gen-tcr}) and~(\ref{eq:gen-qje}), 
provide the robust theoretical basis necessary to do that, as evidenced by the exquisite agreement of our numerical simulations with their predictions, cf. Fig.~\ref{fig:results-all}(a,b,e,f).

The framework informed by the approach based on our QFRs is of general applicability to quantum many-body systems. In particular, our protocol to reveal the existence of charges enables one to uncover more general charges than those related to transport properties and governed by the Mazur bound~\cite{Heidrich-Meisner2003,Prosen2011,Prosen2013,Ilievski2017,Heidrich-Meisner2003,
Ilievski2016}, as illustrated by our analysis of the Dicke model.
Ongoing developments in several physical platforms ---e.g., trapped ions~\cite{Benhelm2008pra,Harty2014,An2015,Kienzler2017} and superconducting circuits~\cite{Chen2014,Zou2014}
--- readily enable precise experimental investigations of this model, which will further inform the development of tools to study non-equilibrium quantum dynamics, and its exploitation in practical tasks~\cite{Kosloff2014,Parrondo2015}.
We remark as well the possibility of simultaneously revealing new charges and determining the associated (equilibrium) inverse temperatures borne in our protocol.

An important question that remains open is how to determine the exact form of the charge operators. Indeed, their intimate dependence on the associated Hamiltonian makes it hard to give a general prescription.
Still, we conjecture that a strategy based on analysing the behaviour of different candidate operators under various driving protocols by means of the marginal TCR, Eq.~(\ref{eq:marginal-tcr}), will illuminate the exact form of the charges. This possibility, which lies beyond the scope of this work, will be explored in the future.

These findings will be relevant to fundamental studies on relaxation and thermalisation of quantum systems~\cite{Eisert2015,Campisi2011,Esposito2015}, and to the advance of quantum simulation and quantum probing protocols that exploit QFRs~\cite{Dorner2013,Mazzola2013,Batalhao2014,Johnson2016thermo,Streif2016}. 
Our results extend the foundations of quantum thermodynamics with charges~\cite{Hickey2014,Guryanova2016,YungerHalpern2016,Gogolin2016,Lostaglio2017} to non-equilibrium processes with various (generalised) baths and/or that break one or more conservation laws, while they revert to the standard QFRs in the case that the Hamiltonian is the only conserved quantity ($\numcons=\numcons'=0$) and both processes start at the same inverse temperature, $\beta'=\beta$.
In this case, $\calWBW=-\calWFW= -\beta w$ 
and Eq.~(\ref{eq:gen-tcr}) 
becomes the standard TCR~\cite{Tasaki2000b} with the standard free energy
$F = -\beta^{-1}\ln\partfgge$. In the same conditions, Eq.~(\ref{eq:gen-qje}) recovers the standard QJE~\cite{Tasaki2000a,Kurchan2000,Yukawa2000}.
In addition, if $\numcons=\numcons'$ and $\vecbeta=\vecbeta'$, Eq.~(\ref{eq:gen-tcr}) reduces to the version of the TCR for GGEs derived in Ref.~\cite{Hickey2014} under the assumption of continuity of the charges during the driving protocols.

We further remark that our QFRs do not assume any particular relationship between the sets of observables that commute with $\Hini$ and $\Hfin$; in particular, we do not assume
continuity between these sets~\cite{Hickey2014}; this is important, as frequently even a small perturbation of a Hamiltonian transforms its charges into dynamical quantities.
Thus, our generalised QFRs enable us to study a larger class of non-equilibrium processes, such as cyclic processes that include an intermediate thermalisation step where the system remains isolated and thus equilibrates to a GGE whose generalised temperatures are not fixed beforehand (see Supplementary Note 2 for more details).
This opens the door to studying the thermalisation of an integrable system when perturbed away from integrability~\cite{Kinoshita2006,Langen2015,Tang2017}.
We thus expect our work will contribute to illuminate open fundamental questions on thermalisation, localisation, and ergodicity, especially in the presence of integrals of motion~\cite{Moeckel2008,Trotzky2012,Bertini2015,Ros2015,DAlessio2016}.
%
Besides, due to the enhanced role of fluctuations in small systems, we expect our work will contribute to a better understanding and improved design of new micro- and nano-meter sized devices where the interplay of thermal and quantum effects is paramount~\cite{Pekola2015,Kosloff2014,Bauer2012}, and thus to
address 
questions related to cyclic protocols and the efficiency of quantum heat engines~\cite{Kosloff2014,Parrondo2015},
thermometry of strongly-correlated systems at ultra-low temperatures~\cite{Johnson2016thermo,Correa2015}, 
and novel quantum sensing applications based on quantum information theory and quantum thermodynamics~\cite{
Streif2016,Goold2016,Cosco2017}.
%



In summary, we have derived a set of generalised quantum fluctuation relations relevant to unitary non-equilibrium processes starting from states of the GGE, which correspond to equilibrium states of quantum system with charges.
Based on these generalised QFRs, we have proposed a general method to address the question of identifying all charges relevant to the non-equilibrium dynamics of a quantum system~\cite{Prosen2011,Prosen2013,Ilievski2017,
Ilievski2016}.
We provided robust numerical evidence of the measurable impact that these theoretical findings can have in current studies with quantum simulators; in particular, we highlighted the importance of identifying all conserved charges to obtain unbiased estimates of their equilibrium properties, such as the temperature, through non-equilibrium measurements~\cite{Dorner2013,Mazzola2013,Batalhao2014,Johnson2016thermo}.
In addition, we put forward a scheme 
to determine whether an integral of motion is affected by a class of non-equilibrium processes, without measuring it.


\mysection{Methods} \label{sec:Methods}%
{\small
\mysubsection{Derivation of the generalised QFRs}
To derive the TCR~(\ref{eq:gen-tcr}), let us introduce the shorthand notations
$ \calA_{\vi} = \beta E_{i_0} + \sum_k \beta_k \chargeQ_{k,i_k}$
and
$ \calA'_{\vf} = \beta' E_{f_0}' + \sum_l \beta_l' \chargeQ_{l,f_l}'$, with
$\ket{\vi}=\ket{i_0,\ldots,i_\numcons,\eta}$ and
$\ket{\vf'}=\ket{f_0,\ldots,f_{\numcons'},\eta'}$.
Then, the probability that a realisation of the protocol requires an amount $\calW$ of generalised work~(\ref{eq:gen-work-fw}) reads
\begin{align}
 &\PFW
 (\calW)
 \equiv
 \mathlarger{\sum}_{\vi,\vf} \,
  p_{\vi} \pi_{\vi \to \vf}(\U)
  \delta \! \left[ \calW - \left( \calA' - \calA \right) \right] \:,
 \label{eq:PUgen}
\end{align}
Here $p_{\vi} = \exp(-\calA_{\vi})/\partfgge$, $\partfgge=\sum_{\vi} \exp(-\calA_{\vi})$, is the probability to find the system in state $\ket{\vi}$ in the first projective measurement at $t=0$, and 
$\pi_{\vi \to \vf}(\U) = |\braket{\vf' | \U | \vi}|^2$ is the probability that the system 
initially in state $\ket{\vi}$ is found in state $\ket{\vf'}$ after the protocol $\U$; finally, $\delta(0)=1$ and otherwise $\delta(x)=0$.
The PDF~(\ref{eq:PUgen}) can be rewritten as
\begin{align}
 \PFW
 &(\calW)
 =
 \nonumber \\
 &
 =\mathlarger{\sum}_{\vi,\vf} \,
  \frac{\exp(-\calA_{\vi})}{\partfgge}
  |\braket{\vf' | U | \vi}|^2
  \delta \! \left[ \calW - \left( \calA'_{\vf} - \calA_{\vi} \right) \right]
 \nonumber \\
 &=\mathlarger{\sum}_{\vi,\vf} \,
  \frac{\exp(\calW-\calA'_{\vf})}{\partfgge}
  |\braket{\vf' | U | \vi}|^2
  \delta \! \left[ \calW - \left( \calA'_{\vf} - \calA_{\vi} \right) \right]
 \nonumber \\
 &=\frac{\partfgge'}{\partfgge} 
  \mathlarger{\sum}_{\vi,\vf} \,
  \frac{\exp(\calW -\calA'_{\vf})}{\partfgge'}
  |\braket{\vi | U^{-1} | \vf'}|^2
  \delta \! \left[ \calW + \left( \calA_{\vi} - \calA'_{\vf} \right) \right]
 \nonumber \\ 
 &\equiv
 \frac{\partfgge'}{\partfgge}e^{\calW}
  \PBW(-\calW) \:.
 \label{eq:proof-tcr}
\end{align}
In the first step we substituted $p_{\vi}$ and $\pi_{\vi \to \vf}$, 
in the third step we applied that $\U$ is unitary,
and in the last step we identified the PDF corresponding to the time-reversed process.
This completes the proof of Eq.~(\ref{eq:gen-tcr}), from which the generalised QJE~(\ref{eq:gen-qje}) follows as discussed in the main text.
The PDFs of standard (dimensionful) work, $w=\Efin'-\Eini$, shown in Fig.~\ref{fig:results-all}, are defined analogously, e.g., 
$ \pFW(w) =
 \sum_{\vi,\vf}
  p_{\vi} \pi_{\vi \to \vf}
  \delta \! \left[ w - ( \Efin' - \Eini ) \right] $,
which follows from the standard situation with the same equilibrium conditions at the start of FW and BW processes, $\beta'=\beta$. 
Finally, the marginal TCR~(\ref{eq:marginal-tcr}) is derived in a manner analogous to~(\ref{eq:proof-tcr}) by defining the PDF of marginal work as
$ \PFWm(\calW_m=x)
 \equiv 
 \sum_{\vi,\vf}
  p_{\vi} \pi_{\vi \to \vf} 
  \delta \! \left[ x - ( \calA'^{(m)}_{\vf} - \calA^{(m)}_{\vi} ) \right] $
with
$\calA^{(m)}_{\vi} = \beta\Eini + \sum_{k\neq m}^{\numcons} \beta_k \chargeQkini$
and 
$\calA'^{(m)}_{\vf} = \beta' \Efin' + \sum_{l\neq m}^{\numcons'} \beta_l' \chargeQlfin'$.

We remark that the fundamental assumptions underlying our generalised QFRs are (i) that the state of the system at the start of both FW and BW processes is of the GGE form, and (ii) the the corresponding driving protocols are time-reversed of each other.

\mysubsection{Dicke model with trapped ions}
Consider $N$ ions in an ion trap~\cite{Haffner2008}.
Each ion can be described as a two-level system (qubit) with internal states 
$\{\ket{\uparrow}, \ket{\downarrow} \}$ corresponding to two Zeeman levels 
within the ground $^2S_{1/2}$ electronic state. Their energy splitting can be controlled by an external magnetic field, $B$, as $\hbar\omegaat=\Delta\mu B$, where $\Delta\mu$ is the difference in magnetic moments of the two internal states~\cite{Benhelm2008pra,Harty2014}.
The motional state of the ions in the trap is characterised by $N-1$ collective modes in each direction~\cite{Haffner2008}.
Among these, the centre-of-mass (COM) mode, with eigenfrequency $\omegacom$, is characterised by coupling identically to all ions.
Internal and motional states can be coupled by radiation of frequency $\omega$ close to $\omegaat\pm\omegacom$, the blue (red) motional sideband~\cite{Haffner2008}.

The complete Hamiltonian describing the dynamics of the system reads
$H=H_0 + H_\mathrm{JC} + H_\mathrm{aJC}$
with 
$H_0 = \hbar\omegacom (\hatb^\dagger \hatb + 1/2) +\sum_{l=1}^N \hbar\omegaat \sigma_z^{(l)}$, where $\hatb^\dagger$ and $\hatb$ are the operators creating and annihilating excitations (phonons) in the COM mode and 
$\sigma_z^{(l)}
= \ket{\uparrow}_{(l)}\bra{\uparrow}
- \ket{\downarrow}_{(l)}\bra{\downarrow}$
is the Pauli $z$ operator for ion $l=1,\ldots,N$.
The coupling between the ions' internal state and the COM mode mediated by radiation is given by the Jaynes-Cummings (JC) and anti-JC Hamiltonians, 
$H_\mathrm{JC}
 =\sum_{l=1}^N 
 \left( \hatb \sigma_+^{(l)} + \hatb^\dagger\sigma_-^{(l)} \right)
 \hbar\omegarsb/2$,
and
$H_\mathrm{aJC}
 =\sum_{l=1}^N 
 \left( \hatb^\dagger \sigma_+^{(l)} + \hatb\sigma_-^{(l)} \right)
 \hbar\omegabsb/2$,
with the raising (lowering) operators
$\sigma_+^{(l)} = \ket{\uparrow}_{(l)}\bra{\downarrow}$
and $\sigma_-^{(l)}=[\sigma_+^{(l)}]^\dagger$.
$\omegarsbbsb$ is the Rabi frequency of the blue (red) motional sideband~\cite{Haffner2008}.

The internal quantum state of a single ion can be mapped onto an effective spin-1/2 system.
The full quantum state of the ions in the trap can then be expressed in the basis $\ket{J,J_z,n} = \ket{J,J_z}\otimes \ket{n}$, where $\ket{n}$ is the Fock state with $n=0,1,2,\ldots$ excitations in the COM mode, and $|J,J_z\rangle$ is an eigenstate of the collective Schwinger pseudo-spins $\hatJ_s = \sum_{l=1}^N \sigma_s^{(l)}$ ($s=z,+,-$), with $J=N/2$ and $J_z=-J,-J+1,\ldots,J$.
Using the collective pseudo-spins and dropping constant terms, the Hamiltonian $H$ can be conveniently rewritten as Eq.~(\ref{eq:Dicke}), with the coupling parameters given in terms of the Rabi frequencies within $H_\mathrm{JC}$ and $H_\mathrm{aJC}$ by
$\lambdaDicke=(\omegarsb+\omegabsb)/2$
and
$\alpha=\omegabsb/(\omegarsb+\omegabsb)$.

\mysubsection{Numerical calculations}
We solve the time evolution of the system with $N=7$ ions
by exact propagation with the full interacting Hamiltonian expressed in the basis of eigenstates $\ket{J,J_z,n}$ with $J=7/2$, $J_z=-7/2,\ldots, 7/2$,
and $n=0,1,\ldots,n_{\mathrm{max}}$. We have verified that a maximum phonon occupation $n_{\mathrm{max}}=800$ is sufficient to faithfully simulate the evolution for the timescales of interest.
To simulate the two-projective-measurement protocol, we proceed in the following way. We start the process in a given eigenstate of the system, with definite eigenvalues of the Hamiltonian, $\Ham$, and the conserved charge, $\hQ$, $\ket{E_n, \chargeQ_m }$. Then, we perform a sudden quench to the intermediate stage and, then, another quench to the final Hamiltonian. We calculate the probability of each transition $\ket{ E_n, \chargeQ_m } \rightarrow \ket{ E_p', \chargeQ_r'}$, involving a work $w=E_p' - E_n$ and a change in the conserved charge, $w_{\chargeQ}= \chargeQ_r' - \chargeQ_m$; this probability is
$P(w,w_\chargeQ) = \left| \braket{E_p',\chargeQ_r'|E_n,\chargeQ_m } \right|^2$.
From this result, we obtain the marginal probabilities for the work, $w$, and the change in the charge, $w_\chargeQ$; both values provide us the generalised work required by the transition, Eqs.~(\ref{eq:gen-work-fw})-(\ref{eq:gen-work-bw})~\cite{Hickey2014,Guryanova2016,YungerHalpern2016}. We repeat the same calculations for every eigenstate of the initial system, obtaining the corresponding marginal probability distributions. The final results follow by averaging the different initial states with the probability distribution given by the GGE, with the corresponding temperatures $\beta$ and $\betaQ$. Note that this procedure is totally equivalent to averaging over a large number of realisations consisting in: first selecting randomly an initial eigenstate $\ket{E_n, \chargeQ_m }$, with the probability distribution given by the GGE (simulating the first projective measurement); and second, selecting randomly the final state as an eigenstate of the final Hamiltonian, with a probability distribution given by $\left| \braket{E_p',\chargeQ_r'|E_n,\chargeQ_m } \right|^2$ (simulating the second projective measurement).
Moreover, this numerical procedure is in direct analogy with the implementation of the filtering method in~\cite{An2015} to project the initial state onto a given eigenstate of the system.

In all the simulations shown, we use $\lambdaDicke(t<0)=2\enUnit$, $\lambdaDicke(0<t<\tfin)=3\enUnit$ and $\lambdaDicke(\tfin)=\enUnit$.
This choice entails that the coupling constant in the intermediate stage is above the critical coupling, $\lambdaDicke_{\mathrm{cr}} = \sqrt{\omega_{\text{COM}}\omega_{\text{at}}}/2 \sim 2.74\enUnit$, for the transition from normal to super-radiant behaviour of the Dicke model (see 
Supplementary Note 3). Indeed, we have checked that the majority of the populated levels at the end of the protocol lays in the chaotic regime. Thus, we expect an effective breakdown of the conservation of $\hQ$, and a complete thermalisation for sufficiently large $\tfin$~\cite{Neill2016}.
} 


\medskip
\mysubsection{Data availability}
The data that support the findings of this study are available
from the corresponding author on request.

\medskip
\mysection{Acknowledgements}
{\small
 We acknowledge useful discussions with B. Bu\v{c}a, J.\ Dukelsky, J.\ J.\ Garc{\'\i}a-Ripoll, K.\ Hovhannisyan, D.\ Jennings, D.\ Lucas, and K.\ Thirumalai.
 This work was supported by the EU H2020 Collaborative project QuProCS (Grant Agreement No.\ 641277),
 EU Seventh Framework Programme QMAC (Grant Agreement No.\ 319286), 
 EPSRC Grant No.\ EP/P01058X/1,
 Spain's MINECO/FEDER Grants
 Nos.\ FIS2015-70856-P, 
 FIS2012-35316, FIS2015-63770-P, 
 and FIS2014-61633-EXPLORA, 
 and CAM research consortium QUITEMAD+ (S2013/ICE-2801).
 DJ thanks the Graduate School of Excellence Material Science in Mainz for hospitality during part of this work.
}

\medskip
\mysection{Author contributions}
{\small
 JMP envisioned the project, derived analytically the generalised QFRs, and devised the implementation of the Dicke model with trapped ions.
 AR performed all numerical simulations.
 JMP, AM, RAM, and DJ contributed to discussion and interpretation of the results, and to the final manuscript.
}
%
%


\renewcommand{\refname}{\normalfont\selectfont\normalsize References}

%

\clearpage


\onecolumngrid

\begin{center}
\textbf{\large Supplementary Information: \\ \mytitle}

J. Mur-Petit, A. Rela\~no, R. A. Molina, and D. Jaksch
\end{center}

\setcounter{equation}{0}
\setcounter{figure}{0}
\setcounter{section}{0}
\setcounter{table}{0}
\setcounter{page}{1}

\renewcommand\theequation{S.\arabic{equation}}

\renewcommand\thefigure{\arabic{figure}}
\renewcommand{\figurename}{Supplementary Figure}

\newcommand{\secname}{Supplementary Note}
\renewcommand\thesection{\secname ~\arabic{section}}

\newcommand{\mySupDisc}{\@startsection {section}{1}{\z@}%
                                   {-3.5ex \@plus -1ex \@minus -.2ex}%
                                   {2.3ex \@plus.2ex}%
                                   {\normalfont\normalsize\bfseries}}
\newcommand{\mySupNote}{\@startsubsubsection {subsubsection}{1}{\z@}%
                                   {-3.5ex \@plus -1ex \@minus -.2ex}%
                                   {2.3ex \@plus.2ex}%
                                   {\normalfont\normalsize\bfseries}}

\makeatletter
\renewcommand\section{\@startsection {section}{1}{\z@}%
                                   {-3.5ex \@plus -1ex \@minus -.2ex}%
                                   {2.3ex \@plus.2ex}%
                                   {\normalfont\normalsize\bfseries}}
\makeatother


\twocolumngrid

\vspace*{1em}%
\noindent\textbf{\textsc{Supplementary Note 1}} \newline
\noindent\textbf{Equilibration and thermalisation in systems with charges}
\vspace*{1em}

The main result of our paper is the derivation of a set of quantum fluctuation relations that apply to non-equilibrium processes in system whose initial state is of the form of a GGE density matrix (Eq.~(\ref{eq:rhoGGE}) in main text). Such a density matrix describes an equilibrium thermal state for a system with charges, and so it is important to discuss if and when one can expect a quantum system to be able to relax to such a state, at least in the sense that expectation values calculated with a GGE density matrix constitute reliable predictions of the system's physical properties.
The study of equilibration and thermalisation in isolated quantum systems is a very active research field, both theoretically and experimentally. In this Supplementary Discussion we review its most relevant results and how they relate to the main part of our work.

As in the main text, we will be concerned with the evolution of a quantum system starting from a certain initial state ---which we do not require now to be an equilibrium state--- in absence of external perturbations.
If the initial state is pure, $\ket{ \psi(0) }$, the time-evolved state remains always pure, $| \psi(t) \rangle$, and therefore nothing like a relaxation to a (mixed) equilibrium state occurs.
However, 
under almost any circumstances,
the density matrix corresponding to the actual time-evolved state, $\rho(t) = \ket{ \psi(t) }\bra{\psi(t)}$, remains almost always close to an effective equilibrium state
$\rho_{\text{eq}} = \lim_{T \rightarrow \infty} (1/T) \int_0^T dt \, \rho(t)$: the amount of time during which $\rho(t)$ is far from $\rho_{\text{eq}}$ is negligible \cite{SM:Gogolin2016}.
As a consequence, the long-time average 
of any physical observable $\hat{\mathcal O}$ coincides with the statistical average over the former effective equilibrium state,
$\langle \hat{\mathcal O} \rangle_{\text{eq}} = \text{Tr} \left[ \rho_{\text{eq}} \hat{\mathcal O} \right]$.
This result is mathematically proven under very general assumptions \cite{
SM:Neumann1929,SM:Goldstein2009,SM:Short2011,SM:Short2012,SM:Reimann2012}, so it is expected to hold in any experiment. Unfortunately, the proof says nothing about the shape of $\rho_{\text{eq}}$; in particular, it does not link $\rho_{\text{eq}}$ with the usual statistical ensembles, like the Generalised Gibbs Ensemble, $\rho_{\text{GGE}}$. Furthermore, it is straightforward to see that, if the spectrum of the system is non-degenerate, $E_m \neq E_n$, $\forall n \neq m$, this equilibrium state reads,
\begin{equation}
\rho_{\text{eq}} = \sum_n \left |C_n \right|^2 \left| n \rangle \langle n \right|,
\label{sm:eq:rho_eq}
\end{equation}
where $| n \rangle$ represents the eigenstate of the system with energy $E_n$; the coefficients $C_n$ depend on the initial state, $C_n = \langle \psi(0) |  n \rangle$, and $| \psi(0) \rangle$ is the initial state itself. In other words, {\em the equilibrium state stores a lot of information about the initial state}, encoded in the coefficients $\left| C_n \right|^2$. This fact is compatible neither with standard thermodynamics (according to which the expectation value of the energy, $E$, is enough to properly describe the equilibrium state), nor with the predictions of the GGE (which establishes that the equilibrium state depends on the expectation values of a few macroscopic observables: a small number of conserved charges, $\hat{M}_k$). %
Indeed, Supplementary Eq.~(\ref{sm:eq:rho_eq}) is compatible with GGE only if the number of conserved charges is equal to the dimension of the phase space, and $\hat{M}_k = \ket{k}\bra{k}$. 

This paradox is solved by the eigenstate thermalisation hypothesis (ETH), at least in highly chaotic quantum systems~\cite{SM:Rigol2007,SM:Rigol2008}. In this kind of systems, expectation values of {\em physical observables}, $\hat{\mathcal O}$, change smoothly with energy, i.e.,
\begin{align}
{\mathcal O}_{n,n}
&
\equiv \langle n | \hat{\mathcal O} | n \rangle
= f(E)
\nonumber \\
&
\approx f(E_n) + \frac{1}{2}\frac{\partial^2 f}{\partial E^2} \left(E_n - E \right)^2 + O \left(E_n - E \right)^3
\end{align}
with the linear term vanishing by symmetry \cite{SM:Jensen1985,SM:Deutsch1991,SM:Srednicki1994,SM:Tasaki1998,SM:Rigol2008}.
As a consequence, the expectation values of such observables in the equilibrium ensemble read
\begin{equation}
\langle \hat{\mathcal O} \rangle_{\text{eq}} = \sum_n \left| C_n \right|^2 {\mathcal O}_{n,n} = \sum_n \left| C_n \right|^2 f(E_n).
\end{equation}
Let us now consider an initial state for which the values of the coefficients $\left| C_n \right|^2$ are peaked around an energy $E$, with a width $\sigma_E$ characterising the spread of the distribution.
If the distribution is `narrow', i.e., $\sigma_E^2 \left( \partial^2 f / \partial E^2 \right) \ll f(E)$, we can safely say that
\begin{equation}
\langle \hat{\mathcal O} \rangle_{\text{eq}} = \sum_n \left| C_n \right|^2 f(E) = f(E), 
\label{sm:eq:eth}
\end{equation}
independently of the actual values of the coefficients $C_n$~\cite{SM:Srednicki1996}.
In other words, according to the ETH, the expectation value $\langle \hat{\mathcal O} \rangle$ is the same over any sufficiently narrow distribution. Hence, highly chaotic quantum systems are expected to thermalise independently of the initial condition, $\ket{ \psi(0) }$, at least for {\em physical observables}.

Two important caveats concern this result. First, there is not a clear definition of {\em physical observables} to which Supplementary Eq.~(\ref{sm:eq:eth}) applies. 
For example, it is quite obvious that projectors over the eigenstates,
$\hat{\mathcal O}_j = \ket{ j } \bra{ j }$, do not satisfy the conditions of the ETH. Furthermore, the condition $\sigma_E^2 \left( \partial^2 f / \partial E^2 \right) \ll f(E)$ strongly depends on the shape of the observable; one can easily find operators $\hat{\mathcal O}$ not satisfying it given a particular width of the energy distribution, $\sigma_E^2$.
Second, if a quantum system has a number of charges, the ETH does not hold either. On the contrary, the expectation value of an observable in a particular eigenstate largely depends on the quantum numbers determined by the charges. So, if two consecutive energy levels, say $\ket{n}$ and $\ket{n+1}$, have different values of these quantum numbers, we can generically expect that ${\mathcal O}_{n, n}$ to be notably different from ${\mathcal O}_{n+1, n+1}$.

The GGE is introduced to solve the second caveat. If a quantum system has a number of relevant charges, it is conjectured that 
a GGE density matrix properly describes the expectation values of {\em physical} observables in the equilibrium states. Unfortunately, as discussed in the main text, it is not easy to establish how many charges are required to build the GGE for a generic system. (As we have pointed above, if this number equals the dimension of the Hilbert space, the corresponding GGE exactly reproduces the equilibrium state $\rho_{\text{eq}}$.)
It is possible to find certain initial conditions for which a small number of charges is enough to properly describe equilibrium states, whereas the actual time-evolved state fluctuates around a totally different equilibrium state, if the experiment starts from another initial state. A generalisation to the ETH for these systems, called generalised eigenstate thermalisation hypothesis, has been proposed, but it is not based in so solid grounds as the ETH~\cite{SM:DAlessio2016}.
In fact, an important number of exceptions to its predictions for the XXZ model were reported (see e.g. \cite{SM:Pozsgay2014,SM:Wouters2014,SM:Goldstein2014}).
While these particular difficulties with the XXZ model have been nowadays resolved by the the identification of relevant quasi-local charges~\cite{SM:Prosen2011,SM:Prosen2013,SM:Ilievski2015,SM:Ilievski2017}, this episode points to the limitations of the generalised ETH.

Summarising, even though there exists a mechanism leading to thermalisation in highly chaotic systems, it does apply neither to all kind of observables, nor to systems with conserved charges. So, the choice of the observables used to test whether a quantum system is properly thermalised, or if all the relevant conserved charges are taken into account, is a tricky task that can lead to misleading conclusions. In this work we propose an alternative method based in generalised quantum fluctuation theorems, not depending on particular observables.

\vspace*{1em}%
\noindent\textbf{\textsc{Supplementary Note 2}} \newline
\noindent\textbf{Physical content of the QFRs}
\vspace*{1em}

In the main part of this paper, we have derived new QFRs that relate the fluctuations when a quantum systems is subject to two non-equilibrium processes (labelled `forward' and `backward') starting from equilibrium states with a (possibly) different number of charges, and (possibly) different generalised temperatures, $\vecbeta \neq \vecbeta'$. As standard QFRs are concerned with the situation that the initial state of the FW and BW processes is at the same inverse temperature, we clarify here the physical content derived from considering this more general situation.

The link that Eqs.~(\ref{eq:gen-tcr}) and~(\ref{eq:gen-qje}) in the main text establish between
different GGE states builds on the definitions of the quantities $\calWFW,\calWBW$.
Physically, we can associate a state function to each GGE state; e.g., for the state at the start of the FW process, we define $\calA(\vecbeta) = \beta \eval{\Ham} + \sum_k \beta_k \eval{\hQ_k}$, with $\eval{\cdots}=\Tr[\rG(\vecbeta)\cdots]$, cf.~\cite{SM:Guryanova2016,SM:YungerHalpern2016}.
It is straightforward to see that in absence of charges, $\calA$ reduces to the system's energy multiplied by its inverse temperature.
The quantity $\calWFW$ (respectively, $\calWBW$) then measures how much this state function changes as the process $\U$ (resp. $\U^{-1}$) takes the system away from the initial state, much as standard work measures how much energy is pumped into the system in a non-equilibrium process.
Generally the system at the end of each process is out of equilibrium, and it is not possible to associate to it a definite value of a state function.
However, the microscopic reversibility of the laws of physics, embodied here in the unitarity of $\U$, allows to establish robust connections, in the form of
Eqs.~(\ref{eq:gen-tcr}) and~(\ref{eq:gen-qje}),
between the out-of-equilibrium fluctuations of the energy and charges of an equilibrium state of $\Hini$ with those of any equilibrium state of $\Hfin$.

\vspace*{1em}%
\noindent\textit{TCR and cyclic processes}
\vspace*{1em}

As the TCR measures the statistical asymmetry in the work PDFs of the FW and BW processes, an exact knowledge of both $\vecbeta$ and $\vecbeta'$ is required.
The standard TCR and its generalisation in \cite{SM:Hickey2014} are derived assuming equal baths at the start of both processes. Our formalism makes it possible instead to study more general processes, including:

\begin{itemize}

\item[{\em (i)}] Processes in which the system is connected to equal baths before the forward and the backward parts. They constitute the natural generalisation of the standard TCR; the number of conserved charges and the corresponding temperatures are the same in both initial states. Hence, the physical interpretation of the generalised TCR is straightforward. The mechanical generalised work required to complete the forward part of the process can be defined as
\begin{equation}
W \equiv {\mathcal W}/\beta = w + \sum_{k=1}^{\mathcal N} \frac{\beta_k}{\beta} \Delta M_k,
\end{equation}
where $w$ is the mechanical work. So, $W$ can be understood as a generalised mechanical energy. If the process is done slowly enough, $W = \Delta {\mathcal F}/\beta$, and all of these generalised energy can be recovered in the backward part of the process.
Otherwise, a part of the generalised mechanical energy will be dissipated into the final baths by a flux of generalised heat, ${\mathcal Q} = W - \Delta {\mathcal F}/\beta$, and this flux is linked to the entropy produced by the process, $\Delta S = \beta Q$ (we take $k_B = 1$). After the backward part of the process, when the cycle is completed, $\Delta {\mathcal F} = 0$, and the entropy produced is sent to the rest of the universe.

\item[{\em (ii)}] Processes in which the system is connected to different baths before the forward and the backward parts, including the case of baths that can have a different number of charges. In this case,  the interpretation in terms of the flux of generalised heat and the  increase of entropy is not so clean, since there is no single  temperature $\beta$. However, the magnitude ${\mathcal W} - \Delta {\mathcal F}$ still represents a measure of the irreversibility of  the process. Exactly as in the former case, when the cycle is  completed $\Delta {\mathcal F} = 0$, and the entropy produced is  sent to the rest of the universe.

\item[{\em (iii)}] Processes completed in isolation. In this case, the  system starts at a particular initial state, given by the  temperatures $\beta_k$, and relaxes into another particular state  after the forward part of the process, characterised by a new set of temperatures, $\beta'_k$, which are determined by the details of the process. Since no baths are present in this case, the  irreversibility is linked to the non-adiabatic transitions between  energy levels as a consequence of the process. Then, the backward  part of the process starts from the resulting relaxed state. As no  generalised heat can be transferred to external baths, the cycle  ends with $\Delta {\mathcal F} = 0$ only if the process is slow  enough to avoid any non-adiabatic transitions. Hence, the degree of  irreversibility is measured by $\Delta {\mathcal F}$. If $\Delta  {\mathcal F} > 0$, a part of the generalised mechanical work has  been transformed into irreversible microscopic changes in the system.

\end{itemize}

\vspace*{1em}%
\noindent\textit{QJE and our choice of inverse temperatures in the simulations}
\vspace*{1em}

On the other hand, the QJE measures the fluctuations of ${\mathcal W}$ in the forward part of the process, and hence the (eventual) backward part is totally irrelevant. It is thus fundamental to note that the value ${\mathcal F}'$ used to obtain $\Delta {\mathcal F}$ is {\em  not} the actual generalised free energy of the system after the forward part of the protocol, but the generalised equilibrium free energy evaluated with the final Hamiltonian, $\Hfin$, with temperatures $\vecbeta'$ (for a similar discussion concerning the standard QJE see, for example, the section 6.1 of \cite{SM:DAlessio2016}). In other words, the knowledge of the final equilibrium state after the forward part of the protocol, either arising from a connection to external baths, or from internal dynamics in isolation, is {\em not} required to apply QJE. This equality links a dynamical quantity which depends on the details of the protocol, the generalised mechanical work, with an equilibrium magnitude, the generalised free energy, that only depends on the initial equilibrium state, and the initial and final values of the external parameters of the Hamiltonian. The fact that both generalised QJE and TCR are formulated as functions of (possibly) different number of conserved charges, $\numcons$ and $\numcons'$, and (possibly) different temperatures, $\vecbeta$ and $\vecbeta' $, allows us to apply TCR to many different kinds of non-equilibrium processes, but it is irrelevant for the QJE. Hence, the most simple way to test the goodness of QJE in an experimental setup, and to study the existence of missing charges in the corresponding system, as we propose in the main part of the paper, is to take $\beta_k = \beta'_k$ $\forall k$.

\newpage
\noindent\textbf{\textsc{Supplementary Note 3}} \newline
\noindent\textbf{Review of the Dicke model and choice of simulation parameters}
\vspace*{1em}

The Dicke model was formulated over 60 years ago to describe the interaction of an ensemble of $N$ two-level atoms, with internal energy splitting $\hbar\omegaat$, with a monochromatic radiation field of frequency $\omegacom$~\cite{SM:Dicke1954,SM:Garraway2011}. Its main feature is the transition from normal behaviour to super-radiance at a critical coupling $\lambdaDicke_\mathrm{cr}$, entailing a macroscopic population of the atomic excited state and the photon field, even at zero temperature \cite{SM:Hepp1973,SM:Wang1973,SM:Carmichael1973}.
Recent theoretical progress has highlighted its relevance to study excited-state quantum phase transitions \cite{SM:Perez-Fernandez2011}. For the purposes of this work, it is especially interesting the possibility to analyse a transition from integrability to chaos as a function of a single parameter, $\alpha$~\cite{SM:Emary2003prl,SM:Emary2003pre,SM:Relano2016}

In its most general formulation, the Hamiltonian of the model reads
\begin{align}
 \Ham(\lambdaDicke,\alpha)
 &=
 \hbar\omegacom \hatb^{\dagger} \hatb 
 +\hbar\omegaat \hatJ_z
 + \frac{2\hbar\lambdaDicke}{\sqrt{N}} \Big[ 
 	(1-\alpha) \left( \hatJ_+ \hatb + \hatJ_- \hatb^{\dagger} \right)
 \nonumber \\
 &+\alpha \left( \hatJ_+ \hatb^{\dagger} + \hatJ_- \hatb \right)
 \Big] \:.
\end{align}
If $\alpha=0$, the model is fully integrable; the quantity $\hQ = \hatJ + \hatJ_z + \hatb^{\dagger} \hatb$ is conserved. If $\alpha=1$, the model is also fully integrable; in this case, the conserved quantity beyond the Hamiltonian is $\hQ' = \hatJ + \hatJ_z - \hatb^{\dagger} \hatb$. The integrability is broken for $0 < \alpha < 1$, though it has been recently shown that an approximated second integral of motion, specially in the low-energy region, exists even for $\alpha=1/2$ \cite{SM:Relano2016}. Therefore, this model constitutes an ideal choice to test the new QFRs with initial GGE equilibrium states, by controlling the single parameter, $\alpha$.

We can engineer a huge number of protocols, including:
\begin{itemize}
\item[(a)] the exact conservation of the second integral of motion, for example a quench from $\Hini=\Ham(\lambdaDicke_1, \alpha=0)$ to $\Hfin=\Ham(\lambdaDicke_2, \alpha=0)$;
\item[(b)] the change from two different integrals of motions, like a quench from $\Hini=\Ham(\lambdaDicke_1, \alpha=0)$ to $\Hfin=\Ham(\lambdaDicke_2, \alpha=1)$; and
\item[(c)] the transition from integrable to non-integrable regimes, like quenches with $\Hini=\Ham(\lambdaDicke_1,\alpha\in\{0,1\})$ and $\Hfin=\Ham(\lambdaDicke_2,0 < \alpha < 1)$.
\end{itemize}

\begin{figure*}[tbh]
\centering
\includegraphics[width=\textwidth]{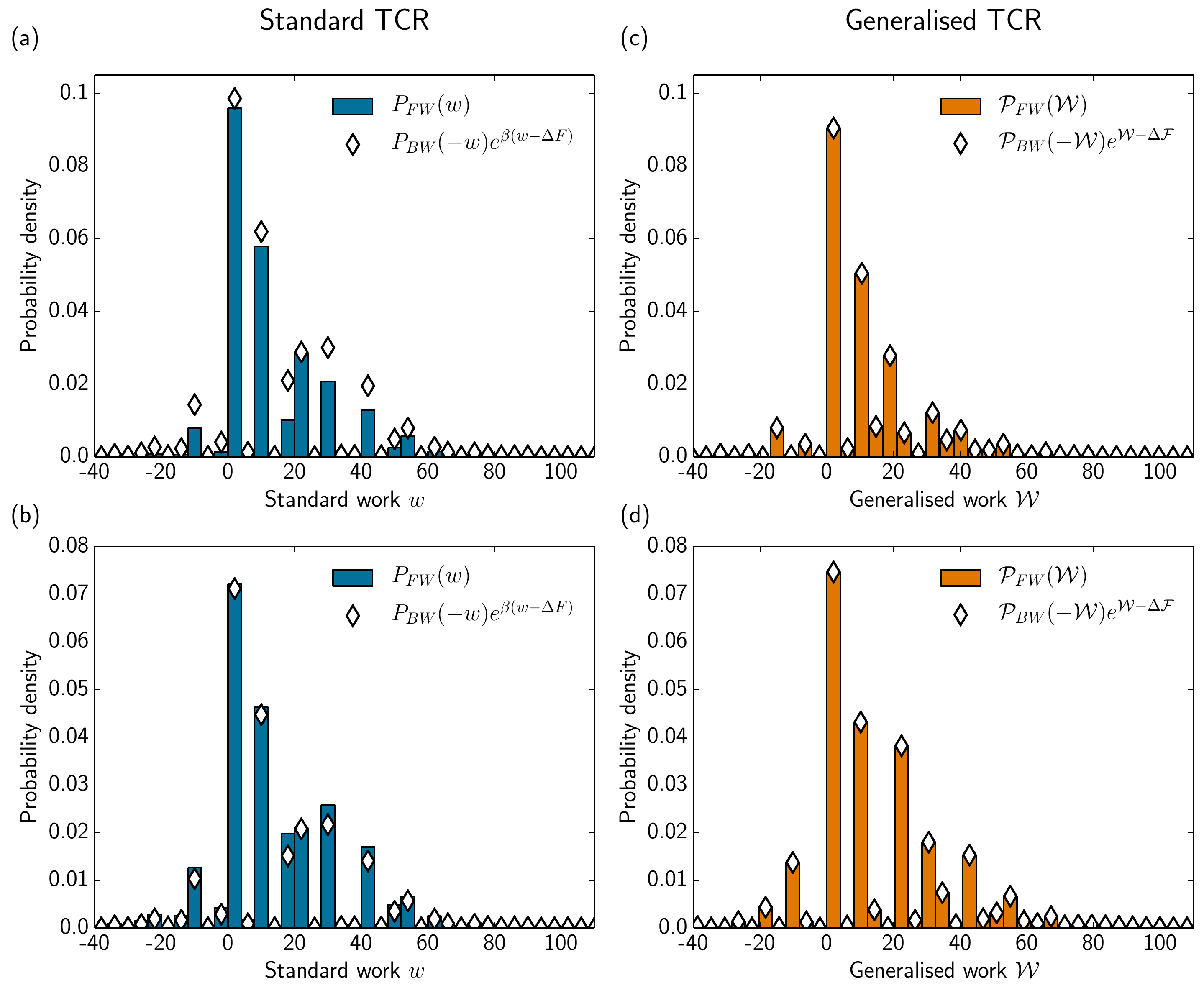}
\caption{ 
 \textbf{Generalised Tasaki-Crooks relation with varying number of charges.}
  Check of the standard and generalised Tasaki-Crooks relations using, respectively,
  standard work (panels a,b) and generalised work (panels c,d) for processes starting with
  $\alpha_\mathrm{FW}(t=0)=0$ and ending at $\alpha_\mathrm{FW}(\tfin)=1/2$, where no additional
  exists. The protocol duration is $\tfin=1.024~\mu\mathrm{s}$,
  and initial equilibrium states are given by:
  (a,b) $(\beta\enUnit,\betaQ\enUnit)=(0.1, 0.3)$ and $\beta'\enUnit=0.1$;
  (c,d) $(\beta\enUnit,\betaQ\enUnit)=(0.1, -0.1)$ and $\beta'\enUnit=0.1$.
 Other parameters:
  $N=7$ ions,
  $\omegacom=3\enUnit$, $\omegaat=10\enUnit$,
  $\gini=2\enUnit$,
  $\gint=3\enUnit$, and
  $\gfin=\enUnit$,
  with $\enUnit=\hbar\times 2\pi$~MHz.
 }
\label{sm:fig:results-alpha12}
\end{figure*}


From this wide variety of possibilities,
in the main part of the text we have studied a protocol of type (a). Specifically, 
we have chosen the following two-quench protocol: $(\lambdaDicke_1=2\enUnit,\alpha=0) \rightarrow (\lambdaDicke_2, \alpha=1/2) \rightarrow (\lambdaDicke_3=\enUnit, \alpha=0)$, allowing the system to remain a time $\tfin$ in the intermediate stage. With this choice, we have the same conserved quantity, $\hQ$, in both the initial and the final stages of the protocol.
This choice has two main features.
First, the initial stages for both the forward and the backward protocols share the same charge $\hQ$. And second, we can control whether $\hQ$ is (approximately) conserved during the protocol, just by changing the duration of the intermediate stage $\tfin$~\cite{SM:Neill2016}.
Thus, the results in the main part of the text highlight both the relevance of charges in work statistics, as well as the fact that the existence of charges at intermediate stages can be accounted for through the generalised QFRs -- and not with the standard ones.

After the system is prepared in the initial state we proceed in the following way. First, we quench the system to the intermediate stage, $(\lambdaDicke_2=3\enUnit, \alpha=1/2)$. We assume that the change in the externals parameters of the Hamiltonian is fast enough to disregard the explicit time dependence of the Hamiltonian; the great majority of the experiments dealing with non-equilibrium processes in small quantum systems are satisfactorily described in this way~\cite{SM:Eisert2015}. Then, we let the system relax in this intermediate stage by evolving with the new values of the parameters for times $0 < t < \tfin$. As discussed above, the value of $\tfin$ is critical to determine whether $\hQ$ is approximately conserved or not: the smaller the value of $\tfin$, the better the (approximate) conservation of $\hQ$ throughout the whole process. Finally, we perform a second quench to the final stage in the same way. 

\newpage
\noindent\textbf{\textsc{Supplementary Note 4}} \newline
\noindent\textbf{Fluctuations with a varying number of charges}
\vspace*{1em}

In the main part of the article, we have discussed numerical results for a protocol involving quenches of the Dicke model, such that the initial Hamiltonian of both FW and BW processes is the same and features one additional charge, $\hQ= \hatJ + \hatJ_z + \hatb^{\dagger} \hatb$.
To illustrate the power of the generalised QFRs to deal with situations where the total number of charges of the system changes as a result of the protocol, we show here results of a single-quench protocol of type (c).
Specifically, we consider the quench $(\lambdaDicke=2\enUnit,\alpha=0) \rightarrow (\lambdaDicke=\enUnit, \alpha=1/2)$.

As $\hQ$ does not commute with $\Hfin$, it does evolve after the quench, and we focus on analysing the PDFs of work and generalised work for after a fixed evolution time, $\tfin=1.024 \hbar/\enUnit$.
These results are shown in Supplementary Fig.~\ref{sm:fig:results-alpha12}(a,b), which correspond to an initial state of the FW process with 
$(\beta,\betaQ)\enUnit = (0.1, 0.3)$ 
or
$(\beta,\betaQ)\enUnit = (0.1, -0.1)$, respectively
(for simplicity, we take the initial state of the BW process to be given by $\beta'=\beta$
as well); these figures are in complete analogy to those in Fig.~2(c-f) of the main text.
Again, we observe that the PDFs of generalised work for the FW and BW processes fulfil the
generalised TCR, Eq.~(\ref{eq:gen-tcr}), 
while the PDFs of standard work noticeably disagree with the predictions from the standard TCR.

Interestingly, for these simulations, we observe how the initial constraint on the allowed values of $\hQ$, is reflected even after the quench ---i.e., when $\hQ$ no longer commutes with the Hamiltonian--- in the PDFs of both generalised and standard work through vanishing probabilities for those values of, respectively, $\calW$ and $w$ that would relate to eigenstate transitions $\ket{E_n,M_m} \to \ket{E_p'}$ with initial states incompatible with the $\hQ$-constraint.


\begin{thebibliography}{70}%
\bibitem{Schilpp2000} P. A. Schilpp, ed., \textit{Albert Einstein: Philosopher-Scientist}, Vol. 7, The Library of Living Philosophers (Open Court Publishing, La Salle, IL, 2000),
pp. 32--33.
\bibitem{Carnot1824} S. Carnot, \textit{R\'eflexions sur la puissance motrice du feu et sur les machines propres \`a d\'evelopper cette puissance} (Bachelier, Paris, 1824).
\bibitem{Liphardt2002} J. Liphardt, S. Dumont, S. B. Smith, I. Tinoco, and C. Bustamante, ``Equilibrium information from nonequilibrium measurements in an experimental test of Jarzynski's equality'', Science \textbf{296}, 1832--1835 (2002)
\href{https://doi.org/10.1126/science.1071152}{[DOI: 10.1126/science.1071152]}.

\bibitem{Ritort2008} F. Ritort, ``Nonequilibrium Fluctuations in Small Systems: From Physics to Biology'', Adv. Chem. Phys. \textbf{137}, edited by S. A. Rice, 31--123 (2008)
\href{https://doi.org/10.1002/9780470238080.ch2}{[DOI: 10.1002/9780470238080.ch2]}.

\bibitem{Esposito2009} M. Esposito, U. Harbola, and S. Mukamel, ``Nonequilibrium fluctuations, fluctuation theorems, and counting statistics in quantum systems'', Rev. Mod. Phys. \textbf{81}, 1665--1702 (2009)
\href{https://doi.org/10.1103/RevModPhys.81.1665}{[DOI: 10.1103/RevModPhys.81.1665]}.

\bibitem{Scully2003} M. O. Scully, M. S. Zubairy, G. S. Agarwal, and H. Walther, ``Extracting Work from a Single Heat Bath via Vanishing Quantum Coherence'', Science \textbf{299}, 862--864 (2003) \href{https://doi.org/10.1126/science.1078955}{[DOI: 10.1126/science.1078955]}.

\bibitem{Abah2014} O. Abah and E. Lutz, ``Efficiency of heat engines coupled to nonequilibrium reservoirs'', EPL (Europhysics Lett.) \textbf{106}, 20001 (2014)
\href{https://doi.org/10.1209/0295-5075/106/20001}{[DOI: 10.1209/0295-5075/106/20001]}.

\bibitem{Rossnagel2014} J. Ro\ss{}nagel, O. Abah, F. Schmidt-Kaler, K. Singer, and E. Lutz, ``Nanoscale heat engine beyond the Carnot limit'', Phys. Rev. Lett. \textbf{112}, 030602 (2014)
\href{https://doi.org/10.1103/PhysRevLett.112.030602}{[DOI: 10.1103/PhysRevLett.112.030602]}.

\bibitem{Hanggi2015} P. H\"anggi and P. Talkner, ``The other QFT'', Nat. Phys. \textbf{11}, 108--110 (2015)
\href{https://doi.org/10.1038/nphys3167}{[DOI: 10.1038/nphys3167]}.

\bibitem{Jarzynski2015commentary} C. Jarzynski, ``Diverse phenomena, common themes'', Nat. Phys. \textbf{11}, 105--107 (2015)
\href{https://doi.org/10.1038/nphys3229}{[DOI: 10.1038/nphys3229]}.

\bibitem{Sutherland2004} B. Sutherland, \textit{Beautiful Models. 70 Years of Exactly Solved Quantum Many-Body Problems} (World Scientific, London, NJ, 2004)
\href{https://doi.org/10.1142/9789812562142_fmatter}{[DOI: 10.1142/9789812562142$\_$fmatter]}.

\bibitem{Essler2005} F. H. L. Essler, H. Frahm, F. G\"ohmann, A. Kl\"umper, and V. E. Korepin, \textit{The One-Dimensional Hubbard Model} (Cambridge University Press, Cambridge, UK, 2005).

\bibitem{Dicke1954} R. H. Dicke, ``Coherence in Spontaneous Radiation Processes'', Phys. Rev. \textbf{93}, 99--110 (1954)
\href{https://doi.org/10.1103/PhysRev.93.99}{[DOI: 10.1103/PhysRev.93.99]}.

\bibitem{Garraway2011} B. M. Garraway, ``The Dicke model in quantum optics: Dicke model revisited.'', Philos. Trans. A. Math. Phys. Eng. Sci. \textbf{369}, 1137--1155 (2011) \href{https://doi.org/10.1098/rsta.2010.0333}{[DOI: 10.1098/rsta.2010.0333]}.

\bibitem{Sutherland1975}  B. Sutherland, ``Model for a multicomponent quantum system'', Phys. Rev. B \textbf{12}, 3795--3805 (1975)
\href{https://doi.org/10.1103/PhysRevB.12.3795}{[DOI: 10.1103/PhysRevB.12.3795]}.

\bibitem{Kinoshita2006} T. Kinoshita, T. Wenger, and D. S. Weiss, ``A quantum Newton's cradle'', Nature \textbf{440}, 900--903 (2006)
\href{https://doi.org/10.1038/nature04693}{[DOI: 10.1038/nature04693]}.

\bibitem{Prosen2011} T. Prosen, ``Open XXZ spin chain: Nonequilibrium steady state and a strict bound on ballistic transport'', Phys. Rev. Lett. \textbf{106}, 217206 (2011)
\href{https://doi.org/10.1103/PhysRevLett.106.217206}{[DOI: 10.1103/PhysRevLett.106.217206]}.

\bibitem{Prosen2013} T. Prosen and E. Ilievski, ``Families of quasilocal conservation laws and quantum spin transport'', Phys. Rev. Lett. \textbf{111}, 057203 (2013) 
\href{https://doi.org/10.1103/PhysRevLett.111.057203}{[DOI: 10.1103/PhysRevLett.111.057203]}.

\bibitem{Ilievski2017} E. Ilievski and J. De Nardis, ``Microscopic Origin of Ideal Conductivity in Integrable Quantum Models'', Phys. Rev. Lett. \textbf{119}, 020602 (2017)
\href{https://doi.org/10.1103/PhysRevLett.119.020602}{[DOI: 10.1103/PhysRevLett.119.020602]}.

\bibitem{Heidrich-Meisner2003} F. Heidrich-Meisner, A. Honecker, D. C. Cabra, and W. Brenig, ``Zero-frequency transport properties of one-dimensional spin-1/2 systems'', Phys. Rev. B \textbf{68}, 134436 (2003) 
\href{https://doi.org/10.1103/PhysRevB.68.134436}{[DOI: 10.1103/PhysRevB.68.134436]}.

\bibitem{Tasaki2000b} H. Tasaki, ``Jarzynski Relations for Quantum Systems and Some Applications''. Preprint available at \url{http://arxiv.org/abs/cond-mat/0009244} (2000).

\bibitem{Tasaki2000a} H. Tasaki, ``Statistical mechanical derivation of the second law of thermodynamics''. Preprint available at \url{http://arxiv.org/abs/cond-mat/0009206} (2000).

\bibitem{Kurchan2000} J. Kurchan, ``A Quantum Fluctuation Theorem''. Preprint available at \url{http://arxiv.org/abs/cond-mat/0007360} (2000).

\bibitem{Yukawa2000} S. Yukawa, ``A Quantum Analogue of the Jarzynski Equality'', J. Phys. Soc. Japan \textbf{69}, 2367--2370 (2000)
\href{https://doi.org/10.1143/JPSJ.69.2367}{[DOI: 10.1143/JPSJ.69.2367]}.

\bibitem{Jaynes1957a} E. T. Jaynes, ``Information Theory and Statistical Mechanics'', Phys. Rev. \textbf{106}, 620--630 (1957) 
\href{https://doi.org/10.1103/PhysRev.106.620}{[DOI: 10.1103/PhysRev.106.620]}.

\bibitem{Rigol2007} M. Rigol, V. Dunjko, V. Yurovsky, and M. Olshanii, ``Relaxation in a completely integrable many-body quantum system: An ab initio study of the dynamics of the highly excited states of 1D lattice hard-core bosons'', Phys. Rev. Lett. \textbf{98}, 050405 (2007)
\href{https://doi.org/10.1103/PhysRevLett.98.050405}{[DOI: 10.1103/PhysRevLett.98.050405]}.

\bibitem{Moreno-Cardoner2007} M. Moreno-Cardoner, J. Mur-Petit, M. Guilleumas, A. Polls, A. Sanpera, and M. Lewenstein, ``Predicting spinor condensate dynamics from simple principles'', Phys. Rev. Lett. \textbf{99}, 020404 (2007) 
\href{https://doi.org/10.1103/PhysRevLett.99.020404}{[DOI: 10.1103/PhysRevLett.99.020404]}.

\bibitem{Talkner2007} P. Talkner, E. Lutz, and P. H\"anggi, ``Fluctuation theorems: work is not an observable.'', Phys. Rev. E \textbf{75}, 050102 (2007) 
\href{https://doi.org/10.1103/PhysRevE.75.050102}{[DOI: 10.1103/PhysRevE.75.050102]}.

\bibitem{Gring2012} M. Gring, M. Kuhnert, T. Langen, T. Kitagawa, B. Rauer, M. Schreitl, I. Mazets, D. A. Smith, E. Demler, and J. Schmiedmayer, ``Relaxation and prethermalization in an isolated quantum system.'', Science \textbf{337}, 1318--1322 (2012) 
\href{https://doi.org/10.1126/science.1224953}{[DOI: 10.1126/science.1224953]}.

\bibitem{Langen2015} T. Langen, S. Erne, R. Geiger, B. Rauer, T. Schweigler, M. Kuhnert, W. Rohringer, I. E. Mazets, T. Gasenzer, and J. Schmiedmayer, ``Experimental observation of a generalized Gibbs ensemble'', Science \textbf{348}, 207--211 (2015) 
\href{https://doi.org/10.1126/science.1257026}{[DOI: 10.1126/science.1257026]}.

\bibitem{Ronzheimer2013} J. P. Ronzheimer, M. Schreiber, S. Braun, S. Hodgman, S. Langer, I. P. McCulloch, F. Heidrich-Meisner, I. Bloch, and U. Schneider, ``Expansion dynamics of interacting bosons in homogeneous lattices in one and two dimensions'', Phys. Rev. Lett. \textbf{110}, 205301 (2013) 
\href{https://doi.org/10.1103/PhysRevLett.110.205301}{[DOI: 10.1103/PhysRevLett.110.205301]}.

\bibitem{Benhelm2008pra} J. Benhelm, G. Kirchmair, C. F. Roos, and R. Blatt, ``Experimental quantum-information processing with $^{43}$Ca$^+$ ions'', Phys. Rev. A \textbf{77}, 062306 (2008) 
\href{https://doi.org/10.1103/PhysRevA.77.062306}{[DOI: 10.1103/PhysRevA.77.062306]}.

\bibitem{Harty2014} T. P. Harty, D. T. C. Allcock, C. J. Ballance, L. Guidoni, H. A. Janacek, N. M. Linke, D. N. Stacey, and D. M. Lucas, ``High-fidelity preparation, gates, memory, and readout of a trapped-ion quantum bit'', Phys. Rev. Lett. \textbf{113}, 220501 (2014) 
\href{https://doi.org/10.1103/PhysRevLett.113.220501}{[DOI: 10.1103/PhysRevLett.113.220501]}.

\bibitem{An2015} S. An, J.-N. Zhang, M. Um, D. Lv, Y. Lu, J. Zhang, Z.-Q. Yin, H. T. Quan, and K. Kim, ``Experimental test of the quantum Jarzynski equality with a trapped-ion system'', Nat. Phys. \textbf{11}, 193--199 (2015) 
\href{https://doi.org/10.1038/nphys3197}{[DOI: 10.1038/nphys3197]}.

\bibitem{Kienzler2017} D. Kienzler, H. Y. Lo, V. Negnevitsky, C. Fl\"uhmann, M. Marinelli, and J. P. Home, ``Quantum Harmonic Oscillator State Control in a Squeezed Fock Basis'', Phys. Rev. Lett. \textbf{119}, 033602 (2017) 
\href{https://doi.org/10.1103/PhysRevLett.119.033602}{[DOI: 10.1103/PhysRevLett.119.033602]}.

\bibitem{Haffner2008} H. H\"affner, C. F. Roos, and R. Blatt, ``Quantum computing with trapped ions'', Phys. Rep. \textbf{469}, 155--203 (2008) 
\href{https://doi.org/10.1016/j.physrep.2008.09.003}{[DOI: 10.1016/j.physrep.2008.09.003]}.

\bibitem{Emary2003prl} C. Emary and T. Brandes, ``Quantum chaos triggered by precursors of a quantum phase transition: the Dicke model.'', Phys. Rev. Lett. \textbf{90}, 044101 (2003) 
\href{https://doi.org/10.1103/PhysRevLett.90.044101}{[DOI: 10.1103/PhysRevLett.90.044101]}.

\bibitem{Emary2003pre} C. Emary and T. Brandes, ``Chaos and the quantum phase transition in the Dicke model'', Phys. Rev. E \textbf{67}, 066203 (2003) 
\href{https://doi.org/10.1103/PhysRevE.67.066203}{[DOI: 10.1103/PhysRevE.67.066203]}.

\bibitem{Relano2016} A. Rela\~no, M. A. Bastarrachea-Magnani, and S. Lerma-Hern\'andez, ``Approximated integrability of the Dicke model'', EPL \textbf{116}, 50005 (2016) 
\href{https://doi.org/10.1209/0295-5075/116/50005}{[DOI: 10.1209/0295-5075/116/50005]}.

\bibitem{Huber2008} G. Huber, F. Schmidt-Kaler, S. Deffner, and E. Lutz, ``Employing Trapped Cold Ions to Verify the Quantum Jarzynski Equality'', Phys. Rev. Lett. \textbf{101}, 070403 (2008) 
\href{https://doi.org/10.1103/PhysRevLett.101.070403}{[DOI: 10.1103/PhysRevLett.101.070403]}.

\bibitem{Dorner2013} R. Dorner, S. R. Clark, L. Heaney, R. Fazio, J. Goold, and V. Vedral, ``Extracting Quantum Work Statistics and Fluctuation Theorems by Single-Qubit Interferometry'', Phys. Rev. Lett. \textbf{110}, 230601 (2013) 
\href{https://doi.org/10.1103/PhysRevLett.110.230601}{[DOI: 10.1103/PhysRevLett.110.230601]}.

\bibitem{Mazzola2013} L. Mazzola, G. De Chiara, and M. Paternostro, ``Measuring the characteristic function of the work distribution.'', Phys. Rev. Lett. \textbf{110}, 230602 (2013) 
\href{https://doi.org/10.1103/PhysRevLett.110.230602}{[DOI: 10.1103/PhysRevLett.110.230602]}.

\bibitem{Batalhao2014} T. B. Batalh\~ao, A. M. Souza, L. Mazzola, R. Auccaise, R. S. Sarthour, I. S. Oliveira, J. Goold, G. De Chiara, M. Paternostro, and R. M. Serra, ``Experimental reconstruction of work distribution and study of fluctuation relations in a closed quantum system.'', Phys. Rev. Lett. \textbf{113}, 140601 (2014) 
\href{https://doi.org/10.1103/PhysRevLett.113.140601}{[DOI: 10.1103/PhysRevLett.113.140601]}.

\bibitem{Hickey2014} J. M. Hickey and S. Genway, ``Fluctuation theorems and the generalized Gibbs ensemble in integrable systems.'', Phys. Rev. E \textbf{90}, 022107 (2014) 
\href{https://doi.org/10.1103/PhysRevE.90.022107}{[DOI: 10.1103/PhysRevE.90.022107]}.

\bibitem{Guryanova2016} Y. Guryanova, S. Popescu, A. J. Short, R. Silva, and P. Skrzypczyk, ``Thermodynamics of quantum systems with multiple conserved quantities'', Nat. Commun. \textbf{7}, 12049 (2016) 
\href{https://doi.org/10.1038/ncomms12049}{[DOI: 10.1038/ncomms12049]}.

\bibitem{YungerHalpern2016} N. Yunger Halpern, P. Faist, J. Oppenheim, and A. Winter, ``Microcanonical and resource-theoretic derivations of the non-Abelian thermal state'', Nat. Commun. \textbf{7}, 12051 (2016) 
\href{https://doi.org/10.1038/ncomms12051}{[DOI: 10.1038/ncomms12051]}.

\bibitem{Gogolin2016}C. Gogolin and J. Eisert, ``Equilibration, thermalisation, and the emergence of statistical mechanics in closed quantum systems.'', Rep. Prog. Phys. \textbf{79}, 056001 (2016) 
\href{https://doi.org/10.1088/0034-4885/79/5/056001}{[DOI: 10.1088/0034-4885/79/5/056001]}.

\bibitem{Lostaglio2017} M. Lostaglio, D. Jennings, and T. Rudolph, ``Thermodynamic resource theories, non-commutativity and maximum entropy principles'', New J. Phys. \textbf{19}, 043008 (2017)
\href{https://doi.org/10.1088/1367-2630/aa617f}{[DOI: 10.1088/1367-2630/aa617f]}.

\bibitem{Johnson2016thermo} T. H. Johnson, F. Cosco, M. T. Mitchison, D. Jaksch, and S. R. Clark, ``Thermometry of ultracold atoms via nonequilibrium work distributions'', Phys. Rev. A \textbf{93}, 053619 (2016) \href{https://doi.org/10.1103/PhysRevA.93.053619}{[DOI: 10.1103/PhysRevA.93.053619]}.

\bibitem{Streif2016} M. Streif, A. Buchleitner, D. Jaksch, and J. Mur-Petit, ``Measuring correlations of cold-atom systems using multiple quantum probes'', Phys. Rev. A \textbf{94}, 053634 (2016) 
\href{https://doi.org/10.1103/PhysRevA.94.053634}{[DOI: 10.1103/PhysRevA.94.053634]}.

\bibitem{Ilievski2016} E. Ilievski, M. Medenjak, T. Prosen, and L. Zadnik, ``Quasilocal charges in integrable lattice systems'', J. Stat. Mech. Theory Exp. \textbf{2016}, 64008 (2016) 
\href{https://doi.org/10.1088/1742-5468/2016/06/064008}{[DOI: 10.1088/1742-5468/2016/06/064008]}.

\bibitem{Chen2014} Y. Chen, P. Roushan, D. Sank, C. Neill, E. Lucero, M. Mariantoni, R. Barends, B. Chiaro, J. Kelly, A. Megrant, J. Y. Mutus, P. J. O'Malley, A. Vainsencher, J. Wenner, T. C. White, Y. Yin, A. N. Cleland, and J. M. Martinis, ``Emulating weak localization using a solid-state quantum circuit'', Nat. Commun. \textbf{5}, 5184 (2014) 
\href{https://doi.org/10.1038/ncomms6184}{[DOI: 10.1038/ncomms6184]}.

\bibitem{Zou2014} L. J. Zou, D. Marcos, S. Diehl, S. Putz, J. Schmiedmayer, J. Majer, and P. Rabl, ``Implementation of the Dicke lattice model in hybrid quantum system arrays'', Phys. Rev. Lett. \textbf{113}, 023603 (2014) 
\href{https://doi.org/10.1103/PhysRevLett.113.023603}{[DOI: 10.1103/PhysRevLett.113.023603]}.

\bibitem{Kosloff2014} R. Kosloff and A. Levy, ``Quantum heat engines and refrigerators: continuous devices'', Annu. Rev. Phys. Chem. \textbf{65}, 365--393 (2014) 
\href{https://doi.org/10.1146/annurev-physchem-040513-103724}{[DOI: 10.1146/annurev-physchem-040513-103724]}.

\bibitem{Parrondo2015} J. M. R. Parrondo, J. M. Horowitz, and T. Sagawa, ``Thermodynamics of information'', Nat. Phys. \textbf{11}, 131--139 (2015)
\href{https://doi.org/10.1038/nphys3230}{[DOI: 10.1038/nphys3230]}.

\bibitem{Eisert2015} J. Eisert, M. Friesdorf, and C. Gogolin, ``Quantum many-body systems out of equilibrium'', Nat. Phys. \textbf{11}, 124--130 (2015) 
\href{https://doi.org/10.1038/nphys3215}{[DOI: 10.1038/nphys3215]}.

\bibitem{Campisi2011} M. Campisi, P. Hänggi, and P. Talkner, ``Colloquium: Quantum fluctuation relations: Foundations and applications'', Rev. Mod. Phys. \textbf{83}, 771--791 (2011) 
\href{https://doi.org/10.1103/RevModPhys.83.771}{[DOI: 10.1103/RevModPhys.83.771]}.

\bibitem{Esposito2015} M. Esposito, M. A. Ochoa, and M. Galperin, ``Nature of heat in strongly coupled open quantum systems'', Phys. Rev. B \textbf{92}, 235440 (2015) 
\href{https://doi.org/10.1103/PhysRevB.92.235440}{[DOI: 10.1103/PhysRevB.92.235440]}.

\bibitem{Tang2017} Y. Tang, W. Kao, K.-Y. Li, S. Seo, K. Mallayya, M. Rigol, S. Gopalakrishnan, and B. L. Lev, ``Thermalization near integrability in a dipolar quantum Newton's cradle''. 
Phys. Rev. X \textbf{8}, 021030 (2018)
\href{https://doi.org/10.1103/PhysRevX.8.021030}{[DOI: 10.1103/PhysRevX.8.021030]}.

\bibitem{Moeckel2008} M. Moeckel and S. Kehrein, ``Interaction Quench in the Hubbard model'', Phys. Rev. Lett. \textbf{100}, 175702 (2008) 
\href{https://doi.org/10.1103/PhysRevLett.100.175702}{[DOI: 10.1103/PhysRevLett.100.175702]}.

\bibitem{Trotzky2012} S. Trotzky, Y.-A. Chen, A. Flesch, I. P. McCulloch, U. Schollw\"ock, J. Eisert, and I. Bloch, ``Probing the relaxation towards equilibrium in an isolated strongly correlated one-dimensional Bose gas'', Nat. Phys. \textbf{8}, 325--330 (2012) 
\href{https://doi.org/10.1038/nphys2232}{[DOI: 10.1038/nphys2232]}.

\bibitem{Bertini2015} B. Bertini, F. H. L. Essler, S. Groha, and N. J. Robinson, ``Prethermalization and Thermalization in Models with Weak Integrability Breaking'', Phys. Rev. Lett. \textbf{115}, 180601 (2015) 
\href{https://doi.org/10.1103/PhysRevLett.115.180601}{[DOI: 10.1103/PhysRevLett.115.180601]}.

\bibitem{Ros2015} V. Ros, M. M\"uller, and A. Scardicchio, ``Integrals of motion in the many-body localized phase'', Nucl. Phys. B \textbf{891}, 420--465 (2015) 
\href{https://doi.org/10.1016/j.nuclphysb.2014.12.014}{[DOI: 10.1016/j.nuclphysb.2014.12.014]}.

\bibitem{DAlessio2016} L. D'Alessio, Y. Kafri, A. Polkovnikov, and M. Rigol, ``From quantum chaos and eigenstate thermalization to statistical mechanics and thermodynamics'', Adv. Phys. \textbf{65}, 239--362 (2016) 
\href{https://doi.org/10.1080/00018732.2016.1198134}{[DOI: 10.1080/00018732.2016.1198134]}.

\bibitem{Pekola2015} J. P. Pekola, ``Towards quantum thermodynamics in electronic circuits'', Nat. Phys. \textbf{11}, 118--123 (2015) 
\href{https://doi.org/10.1038/nphys3169}{[DOI: 10.1038/nphys3169]}.

\bibitem{Bauer2012} G. E. W. Bauer, E. Saitoh, and B. J. van Wees, ``Spin caloritronics'', Nat. Mater. \textbf{11}, 391--399 (2012) 
\href{https://doi.org/10.1038/nmat3301}{[DOI: 10.1038/nmat3301]}.

\bibitem{Correa2015} L. A. Correa, M. Mehboudi, G. Adesso, and A. Sanpera, ``Individual Quantum Probes for Optimal Thermometry'', Phys. Rev. Lett. \textbf{114}, 220405 (2015) 
\href{https://doi.org/10.1103/PhysRevLett.114.220405}{[DOI: 10.1103/PhysRevLett.114.220405]}.

\bibitem{Goold2016} J. Goold, M. Huber, A. Riera, L. del Rio, and P. Skrzypczyk, ``The role of quantum information in thermodynamics -- a topical review'', J. Phys. A Math. Theor. \textbf{49}, 143001 (2016) 
\href{https://doi.org/10.1088/1751-8113/49/14/143001}{[DOI: 10.1088/1751-8113/49/14/143001]}.

\bibitem{Cosco2017} F. Cosco, M. Borrelli, F. Plastina, and S. Maniscalco, ``Momentum-Resolved and Correlations Spectroscopy Using Quantum Probes'', Phys. Rev. A \textbf{95}, 053620 (2017)
\href{https://doi.org/10.1103/PhysRevA.95.053620}{[DOI: 10.1103/PhysRevA.95.053620]}.

\bibitem{Neill2016} C. Neill, P. Roushan, M. Fang, Y. Chen, M. Kolodrubetz, Z. Chen, A. Megrant, R. Barends, B. Campbell, B. Chiaro, A. Dunsworth, E. Jeffrey, J. Kelly, J. Mutus, P. J. J. O'Malley, C. Quintana, D. Sank, A. Vainsencher, J. Wenner, T. C. White, A. Polkovnikov, and J. M. Martinis, ``Ergodic dynamics and thermalization in an isolated quantum system'', Nat. Phys. \textbf{12}, 1037--041 (2016)
\href{https://doi.org/10.1038/NPHYS3830}{[DOI: 10.1038/NPHYS3830]}.
\end{thebibliography}

\begin{thebibliography}{35}%
\bibitem{SM:Gogolin2016} C. Gogolin and J. Eisert, ``Equilibration, thermalisation, and the emergence of statistical mechanics in closed quantum systems.'', Rep. Prog. Phys. \textbf{79}, 056001 (2016) 
\href{https://doi.org/10.1088/0034-4885/79/5/056001}{[DOI: 10.1088/0034-4885/79/5/056001]}.

\bibitem{SM:Neumann1929} J. von Neumann, ``Beweis des Ergodensatzes und des $H$-Theorems in der neuen Mechanik'', Z. Phys. \textbf{57}, 30--70 (1929)
\href{https://doi.org/10.1007/BF01339852}{[DOI: 10.1007/BF01339852]}; English translation: ``Proof of the ergodic theorem and the $H$-Theorem in quantum mechanics'', Eur. Phys. J. H \textbf{35}, 201--237 (2010)
\href{https://doi.org/10.1140/epjh/e2010-00008-5}{[DOI: 10.1140/epjh/e2010-00008-5]}.

\bibitem{SM:Goldstein2009} S. Goldstein, J. L. Lebowitz, C. Mastrodonato, R. Tumulka, and N. Zanghi, ``Normal Typicality and von Neumann's Quantum Ergodic Theorem'', Proc. R. Soc. A Math. Phys. Eng. Sci. \textbf{466}, 3203--3224 (2009)
\href{https://doi.org/10.1098/rspa.2009.0635}{[DOI: 10.1098/rspa.2009.0635]}.

\bibitem{SM:Short2011} A. J. Short, ``Equilibration of quantum systems and subsystems'', New J. Phys. \textbf{13}, 053009 (2011)
\href{https://doi.org/10.1088/1367-2630/13/5/053009}{[DOI: 10.1088/1367-2630/13/5/053009]}.

\bibitem{SM:Short2012} A. J. Short and T. C. Farrelly, ``Quantum equilibration in finite time'', New J. Phys. \textbf{14}, 013063 (2012)
\href{https://doi.org/10.1088/1367-2630/14/1/013063}{[DOI: 10.1088/1367-2630/14/1/013063]}.

\bibitem{SM:Reimann2012} P. Reimann and M. Kastner, ``Equilibration of isolated macroscopic quantum systems'', New J. Phys. \textbf{14}, 043020 (2012)
\href{https://doi.org/10.1088/1367-2630/14/4/043020}{[DOI: 10.1088/1367-2630/14/4/043020]}.

\bibitem{SM:Rigol2007} M. Rigol, V. Dunjko, V. Yurovsky, and M. Olshanii, ``Relaxation in a completely integrable many-body quantum system: An ab initio study of the dynamics of the highly excited states of 1D lattice hard-core bosons'', Phys. Rev. Lett. \textbf{98}, 050405 (2007)
\href{https://doi.org/10.1103/PhysRevLett.98.050405}{[DOI: 10.1103/PhysRevLett.98.050405]}.

\bibitem{SM:Rigol2008} M. Rigol, V. Dunjko, and M. Olshanii, ``Thermalization and its mechanism for generic isolated quantum systems.'', Nature \textbf{452}, 854--858 (2008)
\href{https://doi.org/10.1038/nature06838}{[DOI: 10.1038/nature06838]}.

\bibitem{SM:Jensen1985} R. V. Jensen and R. Shankar, ``Statistical behavior in deterministic quantum systems with few degrees of freedom'', Phys. Rev. Lett. \textbf{54}, 1879--1882 (1985) 
\href{https://doi.org/10.1103/PhysRevLett.54.1879}{[DOI: 10.1103/PhysRevLett.54.1879]}.

\bibitem{SM:Deutsch1991} J. M. Deutsch, ``Quantum statistical mechanics in a closed system'', Phys. Rev. A \textbf{43}, 2046--2049 (1991)
\href{https://doi.org/10.1103/PhysRevA.43.2046}{[DOI: 10.1103/PhysRevA.43.2046]}.

\bibitem{SM:Srednicki1994} M. Srednicki, ``Chaos and quantum thermalization'', Phys. Rev. E \textbf{50}, 888--901 (1994)
\href{https://doi.org/10.1103/PhysRevE.50.888}{[DOI: 10.1103/PhysRevE.50.888]}.

\bibitem{SM:Tasaki1998} H. Tasaki, ``From quantum dynamics to the canonical distribution: General picture and a rigorous example'', Phys. Rev. Lett. \textbf{80}, 1373--1376 (1998) 
\href{https://doi.org/10.1103/PhysRevLett.80.1373}{[DOI: 10.1103/PhysRevLett.80.1373]}.

\bibitem{SM:Srednicki1996} M. Srednicki, ``Thermal fluctuations in quantized chaotic systems'', J. Phys. A. Math. Gen. \textbf{29}, L75--L79 (1996)
\href{https://doi.org/10.1088/0305-4470/29/4/003}{[DOI: 10.1088/0305-4470/29/4/003]}.

\bibitem{SM:DAlessio2016} L. D'Alessio, Y. Kafri, A. Polkovnikov, and M. Rigol, ``From quantum chaos and eigenstate thermalization to statistical mechanics and thermodynamics'', Adv. Phys. \textbf{65}, 239--362 (2016)
\href{https://doi.org/10.1080/00018732.2016.1198134}{[DOI: 10.1080/00018732.2016.1198134]}.

\bibitem{SM:Pozsgay2014} B. Pozsgay, M. Mesty\'an, M. A. Werner, M. Kormos, G. Zar\'and, and G. Tak\'acs, ``Correlations after quantum quenches in the XXZ spin chain: failure of the generalized Gibbs ensemble.'', Phys. Rev. Lett. \textbf{113}, 117203 (2014)
\href{https://doi.org/10.1103/PhysRevLett.113.117203}{[DOI: 10.1103/PhysRevLett.113.117203]}.

\bibitem{SM:Wouters2014} B. Wouters, J. De Nardis, M. Brockmann, D. Fioretto, M. Rigol, and J.-S. Caux, ``Quenching the anisotropic Heisenberg chain: exact solution and generalized Gibbs ensemble predictions.'', Phys. Rev. Lett. \textbf{113}, 117202 (2014)
\href{https://doi.org/10.1103/PhysRevLett.113.117202}{[DOI: 10.1103/PhysRevLett.113.117202]}.

\bibitem{SM:Goldstein2014} G. Goldstein and N. Andrei, ``Failure of the local generalized Gibbs ensemble for integrable models with bound states'', Phys. Rev. A \textbf{90}, 043625 (2014) 
\href{https://doi.org/10.1103/PhysRevA.90.043625}{[DOI: 10.1103/PhysRevA.90.043625]}.

\bibitem{SM:Prosen2011} T. Prosen, ``Open XXZ spin chain: Nonequilibrium steady state and a strict bound on ballistic transport'', Phys. Rev. Lett. \textbf{106}, 217206 (2011)
\href{https://doi.org/10.1103/PhysRevLett.106.217206}{[DOI: 10.1103/PhysRevLett.106.217206]}.

\bibitem{SM:Prosen2013} T. Prosen and E. Ilievski, ``Families of quasilocal conservation laws and quantum spin transport'', Phys. Rev. Lett. \textbf{111}, 057203 (2013)
\href{https://doi.org/10.1103/PhysRevLett.111.057203}{[DOI: 10.1103/PhysRevLett.111.057203]}.

\bibitem{SM:Ilievski2015} E. Ilievski, M. Medenjak, and T. Prosen, ``Quasilocal Conserved Operators in the Isotropic Heisenberg Spin-1/2 Chain.'', Phys. Rev. Lett. \textbf{115}, 120601 (2015)
\href{https://doi.org/10.1103/PhysRevLett.115.120601}{[DOI: 10.1103/PhysRevLett.115.120601]}.

\bibitem{SM:Ilievski2017} E. Ilievski and J. De Nardis, ``Microscopic Origin of Ideal Conductivity in Integrable Quantum Models'', Phys. Rev. Lett. \textbf{119}, 020602 (2017)
\href{https://doi.org/10.1103/PhysRevLett.119.020602}{[DOI: 10.1103/PhysRevLett.119.020602]}.

\bibitem{SM:Guryanova2016} Y. Guryanova, S. Popescu, A. J. Short, R. Silva, and P. Skrzypczyk, ``Thermodynamics of quantum systems with multiple conserved quantities'', Nat. Commun. \textbf{7}, 12049 (2016)
\href{https://doi.org/10.1038/ncomms12049}{[DOI: 10.1038/ncomms12049]}.

\bibitem{SM:YungerHalpern2016} N. Yunger Halpern, P. Faist, J. Oppenheim, and A. Winter, ``Microcanonical and resource-theoretic derivations of the non-Abelian thermal state'', Nat. Commun. \textbf{7}, 12051 (2016)
\href{https://doi.org/10.1038/ncomms12051}{[DOI: 10.1038/ncomms12051]}.

\bibitem{SM:Hickey2014} J. M. Hickey and S. Genway, ``Fluctuation theorems and the generalized Gibbs ensemble in integrable systems.'', Phys. Rev. E \textbf{90}, 022107 (2014)
\href{https://doi.org/10.1103/PhysRevE.90.022107}{[DOI: 10.1103/PhysRevE.90.022107]}.

\bibitem{SM:Dicke1954} R. H. Dicke, ``Coherence in Spontaneous Radiation Processes'', Phys. Rev. \textbf{93}, 99--110 (1954)
\href{https://doi.org/10.1103/PhysRev.93.99}{[DOI: 10.1103/PhysRev.93.99]}.

\bibitem{SM:Garraway2011} B. M. Garraway, ``The Dicke model in quantum optics: Dicke model revisited.'', Philos. Trans. A. Math. Phys. Eng. Sci. \textbf{369}, 1137--1155 (2011) 
\href{https://doi.org/10.1098/rsta.2010.0333}{[DOI: 10.1098/rsta.2010.0333]}.

\bibitem{SM:Hepp1973} K. Hepp and E. H. Lieb, ``On the superradiant phase transition for molecules in a quantized radiation field: the Dicke maser model'', Ann. Phys. (N. Y). \textbf{76}, 360--404 (1973) 
\href{https://doi.org/10.1016/0003-4916(73)90039-0}{[DOI: 10.1016/0003-4916(73)90039-0]}.

\bibitem{SM:Wang1973} Y. K. Wang and F. T. Hioe, ``Phase transition in the Dicke model of superradiance'', Phys. Rev. A \textbf{7}, 831--836 (1973)
\href{https://doi.org/10.1103/PhysRevA.7.831}{[DOI: 10.1103/PhysRevA.7.831]}.

\bibitem{SM:Carmichael1973} H. J. Carmichael, C. W. Gardiner, and D. F. Walls, ``Higher order corrections to the Dicke superradiant phase transition'', Phys. Lett. A \textbf{46}, 47--48 (1973) 
\href{https://doi.org/10.1016/0375-9601(73)90679-8}{[DOI: 10.1016/0375-9601(73)90679-8]}.

\bibitem{SM:Perez-Fernandez2011} P. P\'erez-Fern\'andez, P. Cejnar, J. M. Arias, J. Dukelsky, J. E. Garc\'{\i}a-Ramos, and A. Rela\~no, ``Quantum quench influenced by an excited-state phase transition'', Phys. Rev. A \textbf{83}, 033802 (2011)
\href{https://doi.org/10.1103/PhysRevA.83.033802}{[DOI: 10.1103/PhysRevA.83.033802]}.

\bibitem{SM:Emary2003prl} C. Emary and T. Brandes, ``Quantum chaos triggered by precursors of a quantum phase transition: the Dicke model.'', Phys. Rev. Lett. \textbf{90}, 044101 (2003)
\href{https://doi.org/10.1103/PhysRevLett.90.044101}{[DOI: 10.1103/PhysRevLett.90.044101]}.

\bibitem{SM:Emary2003pre} C. Emary and T. Brandes, ``Chaos and the quantum phase transition in the Dicke model'', Phys. Rev. E \textbf{67}, 066203 (2003) 
\href{https://doi.org/10.1103/PhysRevE.67.066203}{[DOI: 10.1103/PhysRevE.67.066203]}.

\bibitem{SM:Relano2016} A. Rela\~no, M. A. Bastarrachea-Magnani, and S. Lerma-Hern\'andez, ``Approximated integrability of the Dicke model'', EPL \textbf{116}, 50005 (2016) \href{https://doi.org/10.1209/0295-5075/116/50005}{[DOI: 10.1209/0295-5075/116/50005]}.

\bibitem{SM:Neill2016} C. Neill, P. Roushan, M. Fang, Y. Chen, M. Kolodrubetz, Z. Chen, A. Megrant, R. Barends, B. Campbell, B. Chiaro, A. Dunsworth, E. Jeffrey, J. Kelly, J. Mutus, P. J. J. O'Malley, C. Quintana, D. Sank, A. Vainsencher, J. Wenner, T. C. White, A. Polkovnikov, and J. M. Martinis, ``Ergodic dynamics and thermalization in an isolated quantum system'', Nat. Phys. \textbf{12}, 1037--1041 (2016)
\href{https://doi.org/10.1038/NPHYS3830}{[DOI: 10.1038/NPHYS3830]}.

\bibitem{SM:Eisert2015} J. Eisert, M. Friesdorf, and C. Gogolin, ``Quantum many-body systems out of equilibrium'', Nat. Phys. \textbf{11}, 124--130 (2015)
\href{https://doi.org/10.1038/nphys3215}{[DOI: 10.1038/nphys3215]}.
\end{thebibliography}
\end{document}